\pdfoutput=1
\documentclass[10pt,twocolumn,twoside]{IEEEtran}
\usepackage[utf8]{inputenc}
\usepackage[pdftex]{graphicx}
\usepackage[cmex10]{amsmath}
\usepackage{mathtools}
\usepackage[]{fixme}
\usepackage{setspace}
\usepackage{tabularx}
\usepackage{booktabs}
\usepackage{blindtext}

\makeatletter
\let\MYcaption\@makecaption
\makeatother
\usepackage[skip=1pt,font=footnotesize]{subcaption}
\makeatletter
\let\@makecaption\MYcaption
\makeatother

\usepackage{cite}	
\makeatletter
	\def\bstctlcite{\@ifnextchar[{\@bstctlcite}{\@bstctlcite[@auxout]}}
	\def\@bstctlcite[#1]#2{\@bsphack
	\@for\@citeb:=#2\do{%
	\edef\@citeb{\expandafter\@firstofone\@citeb}%
	\if@filesw\immediate\write\csname
	#1\endcsname{\string\citation{\@citeb}}\fi}%
	\@esphack}
\makeatother

\newtheorem{theorem}{Theorem}
\newtheorem{lemma}{Lemma}

\usepackage{pgfplots}
\pgfplotsset{compat=1.14}
\usetikzlibrary{plotmarks}
\newlength\figureheight 
\newlength\figurewidth 

\usepackage{bm}
\usepackage{xspace}		
\usepackage{amssymb}	
\usepackage{bbold}		

\newcommand{\dd}{\mathrm{d}}

\newcommand{\shownumber}{\addtocounter{equation}{1}\tag{\theequation}}
\newcommand{\ii}{^{-1}}
\newcommand{\db}{\ensuremath{\,\textrm{dB}}\xspace}
\newcommand{\const}{\ensuremath{\mathrm{const.}}}

\DeclareMathOperator{\argmin}{arg\,min}

\DeclareMathOperator{\diag}{diag}

\DeclareMathOperator{\tr}{tr}

\DeclareMathOperator{\CN}{CN}

\DeclareMathOperator{\im}{Im}

\newcommand{\minn}[1]{\underset{#1}{\min}\;}
\newcommand{\argminn}[1]{\underset{#1}{\argmin}\;}

\newcommand{\limm}[1]{\underset{#1}{\lim}\;}

\newcommand{\h}{^\mathrm{H}}
\newcommand{\T}{^{\mathrm{T}}}

\newcommand{\bbc}{\mathbb{C}}

\newcommand{\calg}{\ensuremath{\mathcal{G}}\xspace}

\newcommand{\call}{\ensuremath{\mathcal{L}}\xspace}
\newcommand{\calm}{\ensuremath{\mathcal{M}}\xspace}

\newcommand{\calo}{\ensuremath{\mathcal{O}}\xspace}

\makeatletter
 \begingroup
 \catcode`\_=\active
 \protected\gdef_{\@ifnextchar|\subtextup\sb}
 \endgroup
 \def\subtextup|#1|{\sb{\textup{#1}}}
 \AtBeginDocument{\catcode`\_=12 \mathcode`\_=32768 }
\makeatother

\newcommand{\balpha}{\ensuremath{\bm{\alpha}}\xspace}

\newcommand{\bgamma}{\ensuremath{\bm{\gamma}}\xspace}
\newcommand{\bGamma}{\ensuremath{\bm{\Gamma}}\xspace}

\newcommand{\btheta}{\ensuremath{\bm{\theta}}\xspace}

\newcommand{\bmu}{\ensuremath{\bm{\mu}}\xspace}

\newcommand{\bSigma}{\ensuremath{\bm{\Sigma}}\xspace}
\newcommand{\btau}{\ensuremath{\bm{\tau}}\xspace}

\newcommand{\bPhi}{\ensuremath{\bm{\Phi}}\xspace}

\newcommand{\bpsi}{\ensuremath{\bm{\psi}}\xspace}
\newcommand{\bPsi}{\ensuremath{\bm{\Psi}}\xspace}

\bmdefine{\Ba}{a}
\newcommand{\ba}{\ensuremath{\Ba}\xspace}
\newcommand{\bA}{\ensuremath{\mathbf{A}}\xspace}
\bmdefine{\Bb}{b}
\newcommand{\bb}{\ensuremath{\Bb}\xspace}
\newcommand{\bB}{\ensuremath{\mathbf{B}}\xspace}
\bmdefine{\Bc}{c}

\newcommand{\bC}{\ensuremath{\mathbf{C}}\xspace}
\bmdefine{\Bd}{d}

\newcommand{\bD}{\ensuremath{\mathbf{D}}\xspace}
\bmdefine{\Be}{e}

\bmdefine{\Bf}{f}

\bmdefine{\Bg}{g}

\bmdefine{\Bh}{h}
\newcommand{\bh}{\ensuremath{\Bh}\xspace}

\bmdefine{\Bi}{i}

\newcommand{\bI}{\ensuremath{\mathbf{I}}\xspace}
\bmdefine{\Bj}{j}

\bmdefine{\Bk}{k}

\bmdefine{\Bl}{l}

\newcommand{\bL}{\ensuremath{\mathbf{L}}\xspace}
\bmdefine{\Bm}{m}

\bmdefine{\Bn}{n}

\bmdefine{\Bo}{o}

\bmdefine{\Bp}{p}

\bmdefine{\Bq}{q}
\newcommand{\bq}{\ensuremath{\Bq}\xspace}

\bmdefine{\Br}{r}
\newcommand{\br}{\ensuremath{\Br}\xspace}

\bmdefine{\Bs}{s}
\newcommand{\bs}{\ensuremath{\Bs}\xspace}

\bmdefine{\Bt}{t}
\newcommand{\bt}{\ensuremath{\Bt}\xspace}
\newcommand{\bT}{\ensuremath{\mathbf{T}}\xspace}
\bmdefine{\Bu}{u}
\newcommand{\bu}{\ensuremath{\Bu}\xspace}

\bmdefine{\Bv}{v}
\newcommand{\bv}{\ensuremath{\Bv}\xspace}

\bmdefine{\Bw}{w}
\newcommand{\bw}{\ensuremath{\Bw}\xspace}

\bmdefine{\Bx}{x}
\newcommand{\bx}{\ensuremath{\Bx}\xspace}

\bmdefine{\By}{y}
\newcommand{\by}{\ensuremath{\By}\xspace}

\bmdefine{\Bz}{z}
\newcommand{\bz}{\ensuremath{\Bz}\xspace}

\usepackage[final,citecolor=black,bookmarks,bookmarksnumbered,colorlinks=true,linkcolor=black]{hyperref}

\begin{document}
\bstctlcite{IEEEexample:BSTcontrol}
\title{Superfast Line Spectral Estimation}
\author{%
	Thomas L. Hansen,
	Bernard H. Fleury, \textit{Senior Member, IEEE},
	and Bhaskar D. Rao, \textit{Fellow, IEEE}
	\thanks{
		T.~L.~Hansen and B.~H.~Fleury are with the Department of Electronic
		Systems, Aalborg University, Aalborg, Denmark. B.~D.~Rao is with the
		Electrical and Computer Engineering department, University of
		California, San Diego.
		
		The work of T. L. Hansen is supported by the Danish Council for
		Independent Research under grant id DFF--4005--00549.
		
		\textcopyright 2018 IEEE. Personal use of this material is permitted.
		Permission from IEEE must be obtained for all other uses, in any
		current or future media, including reprinting/republishing this
		material for advertising or promotional purposes, creating new
		collective works, for resale or redistribution to servers or lists, or
		reuse of any copyrighted component of this work in other works.

		Digital Object Identifier 10.1109/TSP.2018.2807417
		}
}
\markboth{IEEE Transactions on Signal Processing, VOl. xx, No. xx, month,
2018}{Hansen \MakeLowercase{\textit{et al.}}: Superfast Line Spectral Estimation}
\maketitle

\begin{abstract}
	A number of recent works have proposed to solve the line spectral
	estimation problem by applying off-the-grid extensions of sparse
	estimation techniques. These methods are preferable over classical line
	spectral estimation algorithms because they inherently estimate the model
	order. However, they all have computation times which grow at least
	cubically in the problem size, thus limiting their practical applicability
	in cases with large dimensions. To alleviate this issue, we
	propose a low-complexity method for line spectral estimation, which also
	draws on ideas from sparse estimation. Our method is based on a
	Bayesian view of the problem. The signal covariance matrix is shown to
	have Toeplitz structure, allowing superfast Toeplitz inversion to be used.
	We demonstrate that our method achieves estimation accuracy at least as
	good as current methods and that it does so while being orders of
	magnitudes faster.
\end{abstract}

\section{Introduction}

The problem of line spectral estimation (LSE) has received significant
attention in the research community for at least 40 years. The reason is that
many fundamental problems in signal processing can be recast as LSE; examples
include direction of arrival estimation using sensor arrays
\cite{malioutov-sparse,ottersten-analysis}, bearing and range estimation in
synthetic aperture radar \cite{carriere-high}, channel estimation in wireless
communications \cite{bajwa-sparsechannel} and simulation of atomic systems in
molecular dynamics \cite{andrade-application}.

In trying to solve the LSE problem, classical approaches include subspace
methods \cite{kung-statespace} such as MUSIC \cite{schmidt-music} or ESPRIT
\cite{roy-esprit} which estimate the frequencies based on an estimate of the
signal covariance matrix. These approaches must be augmented with a method for
estimation of the model order. Popular choices include generic information
theoretic criteria (e.g. AIC, BIC) or more specialized methods, such as SORTE
\cite{he-sorte} which is based on the eigenvalues of the estimated signal
covariance matrix. Subspace methods typically perform extremely well if the
model order is known, but their estimation accuracy can degrade significantly if
the model order is unknown.

The stochastic maximum likelihood (ML) method is
known to be asymptotically efficient (it attains the Cram{\'e}r-Rao bound as the
problem size tends to infinity) \cite{ottersten-analysis}. Unfortunately it
also requires knowledge of the model order.

Inspired by the ideas of sparse estimation and compressed sensing, many
papers on sparsity-based LSE algorithms have appeared in recent years, e.g.
\cite{malioutov-sparse,hu-compressed}. In particular, the LSE problem is
simplified to a finite sparse reconstruction problem by restricting the
frequencies to a grid. Such methods inherently estimate the model order,
alleviating the issues arising from separate model order and frequency
estimation in classical methods. The granularity of the grid leads to a
non-trivial trade-off between accuracy and computational requirements. To
forego the use of a grid, so-called off-the-grid compressed sensing methods have been
proposed \cite{tang-offthegrid,bhaskar-atomic,candes-superresolution}. These
methods provably recover the frequencies in the noise-free case under a
minimum separation condition. They suffer from prohibitively high
computational requirements even for moderate problem sizes, see Sec
\ref{sec:experiments}.


In \cite{dublanchet-doa,hansen-sam,badiu-valse} a Bayesian view is taken on
the LSE problem. The model used in stochastic ML is extended with a
sparsity-promoting prior on the coefficients of the sinusoid components.
Thereby inherent estimation of the model order is achieved. These algorithms
generally have high estimation accuracy. Their per-iteration computational
complexity is cubic in the number of sinusoidal components, meaning that their
runtime grows rapidly as the number of components increases.

In this work we introduce the Superfast LSE algorithm for solving the
LSE problem in scenarios where the full measurement vector is available
(complete data case).
The modelling and design of the basic algorithm which we present in Sec.
\ref{sec:algorithm} is based upon the ideas in
\cite{dublanchet-doa,hansen-sam,badiu-valse}.
The main novelty resides in the computational aspects of
Superfast LSE. The derived method is based upon several techniques: a so-called
superfast Toeplitz inversion algorithm \cite{ammar-generalized,ammar-numerical}
(thereof the name of our algorithm), low-complexity Capon beamforming
\cite{musicus-fastmlm}, the Gohberg-Semencul formula \cite{gohberg-convolution}
and non-uniform fast Fourier transforms \cite{greengard-nufft, lee-nufft}.
The Superfast LSE algorithm has the following virtues:
It inherently estimates all model parameters such as the noise variance and
model order and it has low per-iteration computational complexity. Specifically it
scales as $\calo(N\log^2N)$ where $N$ is the length of the observed vector. We
show empirically that it converges after a few iterations (typically less
than 20). This means that for large problem sizes our algorithm can have
computation time orders of magnitude lower than that of current methods. It
does so without any penalty in estimation accuracy. Our numerical experiments
show that Superfast LSE has high estimation accuracy across a wide range of
scenarios, being on par with or better than state-of-the-art algorithms.

Synergistically and computationally efficiently combining the steps in
the algorithm might appear easy after the fact. This is however not the case.
Some other LSE algorithms can benefit in terms of computational effort from our
approach, yet not to the extent achieved with the proposed algorithm. For
instance, the computational methods in Sec. \ref{sec:superfast} can be embedded
in VALSE \cite{badiu-valse}. The resulting scheme will have high computational
complexity due to the variational estimation of the posterior on the
frequencies. Note that our algorithm performs on par with VALSE, but at
a significantly reduced computational effort.

For completeness we also present a semifast version of the algorithm which
works when only a subset of entries in the measurement vector are available.
The Semifast LSE algorithm has per-iteration complexity $\calo(N\hat K^2 +
N\log N)$, where $\hat K$ is the number of estimated sinusoids. Algorithms with
similar per-iteration complexity are derived in
\cite{hansen-sam,champagnat-unsupervised, dublanchet-doa,
tipping-fastmarginal}. We have observed that our algorithm
converges in a smaller number of iterations when compared to the algorithm in
\cite{hansen-sam}, thus leading to lower total runtime.


\textit{Outline:}
In Sec. \ref{sec:alg} we present our modelling and algorithm for LSE. Our
low-complexity computational methods are presented in Sec. \ref{sec:superfast}
(complete date case) and \ref{sec:semifast} (incomplete data case). In Sec.
\ref{sec:mmv} the algorithm is extended to the case of multiple measurement
vectors. Numerical experiments are presented in Sec. \ref{sec:experiments} and
conclusions are given in Sec. \ref{sec:conclusions}.

\textit{Notation:}
We write vectors as \ba and matrices as \bA. The $i$th entry of vector \ba is
denoted $a_i$ or $[\ba]_i$; the $i,j$th entry of matrix \bA is denoted
$\bA_{i,j}$. Let \bb be a binary
vector (containing only zeros and ones) of the same dimension as \ba, then
$\ba_\bb$ denotes a vector which contains those entries in \ba where the
corresponding entry in \bb is one. The Hadamard (entrywise) product is denoted
by $\odot$.

\section{An Algorithm for Line Spectral Estimation}
\label{sec:alg}
\label{sec:algorithm}
We now detail the observation model and the specific objective of the LSE
problem. The observation vector $\by\in\bbc^M$ contains time-domain samples and is
given by
\begin{align}
	\by &= \sum_{k=1}^K \bPhi \bpsi(\tilde\theta_k) \tilde\alpha_k + \bw
		= \bPhi\bPsi(\tilde\btheta) \tilde\balpha + \bw,
	\label{model}
\end{align}
where the steering vector function
$\bpsi(\theta_k):[0,1)\rightarrow\bbc^{N\times 1}$ gives a Fourier vector,
i.e., it has $n$th entry $[\bpsi(\theta_k)]_n\triangleq\exp(j2\pi (n-1)
\theta_k)$ for $n=1,\dots,N$. We also define
$\bPsi(\btheta)\triangleq[\bpsi(\theta_1), \cdots,
\bpsi(\theta_{\dim(\btheta)})]$. The measurement matrix $\bPhi\in\bbc^{M\times
N}$ is either the identity matrix ($M=N$, complete
data case) or made of a subset of rows of a diagonal matrix ($M<N$, incomplete data
case). The vector \bw is a white Gaussian noise vector with component variance
$\beta$. The LSE problem is that of recovering the model order $K$ along with
the frequency $\tilde\theta_k\in[0,1]$ and coefficient $\tilde\alpha_k\in\bbc$
of each component $k=1,\ldots,K$.

\subsection{Estimation Model}
The estimation model and inference approach we present in the following are
adaptions of ideas currently available in the literature. We have carefully
combined these ideas to obtain an iterative scheme which can be implemented
with low complexity as described in Secs. \ref{sec:superfast} and
\ref{sec:semifast}, while achieving a performance comparable to that of
state-of-the-art algorithms.

Our algorithm is based on Bayesian inference in an estimation model which
approximates \eqref{model}. Specifically, to enable estimation of the model
order $K$, we follow \cite{dublanchet-doa,badiu-valse} and employ a model with
$K_|max|\ge K$ components%
\footnote{Since we can never expect to estimate more parameters than the number
of observed observations, we select $K_|max|=M$ in our implementation.}.
Each component has an associated activation variable $z_k\in\{0,1\}$ which is
set to 0 or 1 to deactivate or activate it. The activation variables are
collected in the sparse vector \bz. The effective estimated model order is
given by the number of active components.
Based on \eqref{model} we write our estimation model
\begin{align}
	\by = \sum_{k=1}^{K_|max|} \bPhi\bpsi(\theta_k) z_k \alpha_k + \bw
	= \bA(\btheta_\bz)\balpha_\bz + \bw,
	\label{estModel}
\end{align}
where $\theta_k\in [0,1)$ and $\alpha_k\in\bbc$ are frequencies and
coefficients for $k=1,\ldots,K_|max|$ and we have defined
$\bA(\btheta)\triangleq\bPhi\bPsi(\btheta)$.

Due to the Gaussian noise assumption we have
\begin{align}
	p(\by | \balpha, \bz, \btheta; \beta) = \CN(\by; \bA(\btheta_\bz)
	\balpha_\bz, \beta\bI),
\end{align}
where $\CN(\by; \bmu, \bSigma)$ denotes the probability density function of a
circularly symmetric complex normal random variable $\by$ with mean $\bmu$ and
covariance matrix $\bSigma$.
We assume $\beta\in[\varepsilon_\beta,\infty)$, where $\varepsilon_\beta>0$ is an
arbitrarily small constant which guarantees that the likelihood function is
bounded below.
A Bernoulli prior is used to promote deactivation of some of the components:
\begin{align}
	p(\bz;\zeta) = \prod_{k=1}^{K_|max|} \zeta^{z_k} (1-\zeta)^{1-z_k},
	\label{zPrior}
\end{align}
where $\zeta\in[0,1/2]$ is the activation probability. The restriction
$\zeta\le1/2$ ensures that the prior is sparsity inducing. The coefficients are
assumed to be independent zero-mean Gaussian
\begin{align}
	p(\balpha; \bgamma) = \prod_{k=1}^{K_|max|} \CN(\alpha_k; 0,
	\gamma_k),
	\label{alphaPrior}
\end{align}
where $\gamma_k\in[0,\infty)$ is the active-component variance. Sparsity-promoting priors have
previously been used for both basis selection \cite{wipf-sbl} and LSE
\cite{hansen-sam}. The Bernoulli-Gaussian prior structure that we have adopted
above was first introduced in \cite{kormylo-ml} and used for LSE in
\cite{badiu-valse}.


Even though each $\alpha_k$ is modelled as Gaussian in \eqref{alphaPrior}, the
prior specification is significantly more general than that because the
variance of each component is estimated through $\gamma_k$. In the numerical
investigation we demonstrate that our method works well even when the true
density of each coefficient is not Gaussian.

We finally use an independent and identically
distributed (i.i.d.) uniform prior on the entries in \btheta:
\begin{align}
	p(\btheta) &= \prod_{k=1}^{K_|max|} p(\theta_k)
		= \prod_{k=1}^{K_|max|} 1 = 1.
		\label{thetaPrior}
\end{align}
If further prior information about the frequencies is available, it can easily
be incorporated through $p(\btheta)$.

\subsection{Approach}
By integrating the component coefficients we obtain the marginal likelihood
\begin{align*}
	p(\by | \bz, \btheta; \beta, \bgamma) &=
		\int p(\by | \balpha, \bz, \btheta; \beta) p(\balpha; \bgamma)
		\,\dd \balpha \\
			&= \CN(\by; \bm0, \bC)
			\shownumber\label{py}
\end{align*}
with $\bC \triangleq \beta\bI + \bA(\btheta_\bz) \bGamma_\bz \bA\h(\btheta_\bz)$ and
$\bGamma_\bz\triangleq \diag(\bgamma_\bz)$.

Based on the marginal likelihood we can write the objective
\begin{align*}
	\call(\bz,& \zeta, \beta, \btheta, \bgamma) \triangleq
		- \ln p(\bz,\btheta | \by; \beta, \bgamma, \zeta) \\
		&= - \ln p(\by|\bz,\btheta; \beta, \bgamma)
			p(\bz; \zeta) p(\btheta) + \const \\
		&= \ln |\bC| + \by\h\bC\ii\by \\
			& \hspace{6mm}
			- \sum_{k=1}^{K_|max|} \left( z_k\ln\zeta + (1-z_k)\ln(1-\zeta)\right)
			+ \const
			\shownumber\label{obj}
\end{align*}
The variables $(\bz, \btheta)$ and model parameters $(\beta, \bgamma, \zeta)$
are estimated by minimizing \eqref{obj}, i.e., we seek the maximum a-posteriori
(MAP) estimate of $(\bz, \btheta)$ and the ML estimate of $(\beta, \bgamma,
\zeta)$. Our algorithm employs a block-coordinate descent method to find a
local minimum (or saddle point) of \eqref{obj}.

For fixed $\bz$ the first two terms in \eqref{obj} are equal to the objective
function of stochastic ML \cite{ottersten-analysis}, and our approach can
therefore be viewed as stochastic ML extended with a variable model order.

When the above estimates have been computed, the estimated model order is given
by the number of active components, i.e. $\hat K = ||\hat\bz||_0$, and the
entries of $\hat\btheta_{\hat\bz}$ are the estimated frequencies. The
corresponding coefficients $\balpha_{\hat\bz}$ can be estimated as follows. First,
write the posterior of \balpha as
\begin{align*}
	p(\balpha | \by, \hat\bz, \hat\btheta; \hat\beta, \hat\bgamma)
		&\propto
			p(\by | \balpha, \hat\bz, \hat\btheta; \hat\beta)
			p(\balpha; \hat\bgamma) \\
		&\propto
			\CN(\balpha_{\hat\bz} ; \hat\bmu, \hat\bSigma)
			\:\smashoperator{\prod_{\{k:\hat z_k=0\}}}\:
			\CN(\alpha_k; 0, \hat\gamma_k),
			\shownumber \label{palpha}
\end{align*}
where
\begin{align}
	\hat\bmu &\triangleq \hat\beta\ii \hat\bSigma \bA\h(\hat\btheta_{\hat\bz}) \by
		\label{bmu} \\
	\hat\bSigma &\triangleq \left( \hat\beta\ii\bA\h(\hat\btheta_{\hat\bz})\bA(\hat\btheta_{\hat\bz}) +
		\hat\bGamma_{\hat\bz}\ii \right)\ii.
		\label{bSigma}
\end{align}
As expected the posterior of the coefficients corresponding to inactive
components (those for which $\hat z_k=0$) coincides with their prior. These are
not of interest (they are inconsequential in the model \eqref{estModel}) and
integrating them out gives a Gaussian posterior over $\balpha_{\hat\bz}$. If a
point estimate of $\balpha_{\hat\bz}$ is needed, the MAP (which is also the LMMSE)
estimate $\hat\balpha_{\hat\bz}=\hat\bmu$ can be used%
\footnote{Note that for computational convenience we write
$\hat\bmu=\hat\bgamma_{\hat\bz} \odot \bq$, where $\bq$ is defined by \eqref{q}.
See the text following \eqref{betaUpdate}.}.

\subsection{Derivation of Update Equations}
As mentioned, our algorithm is derived as a block-coordinate descent method
applied on $\call$ in \eqref{obj}. The estimates are updated in the following
blocks: $\hat\bz$, $\hat\zeta$, $\hat\beta$ and $(\hat\btheta_{\hat\bz},
\hat\bgamma_{\hat\bz})$.
Each update is guaranteed not to increase $\call$. We note that the frequencies
and variances of inactive components (those for which $\hat z_k=0$)
are not updated, as $\call$ does not depend on these variables.

\subsubsection{Estimation of frequencies and coefficient variances}
\label{sec:thetaUpdate}
Even when all remaining variables are kept fixed, it is not tractable to find
the global minimizer of $\call$ with respect to the vector of active
component frequencies $\btheta_{\hat\bz}$ and variances $\bgamma_{\hat\bz}$.
We therefore resort to a numerical method. Writing only the terms of
\eqref{obj} which depend on $\btheta_{\hat\bz}$, we have
\begin{align*}
	\call(\btheta_{\hat\bz}, \bgamma_{\hat\bz})
	= \ln|\bC| + \by\h\bC\ii\by + \const,
\end{align*}
so we need to solve
$(\hat\btheta_{\hat\bz}, \hat\bgamma_{\hat\bz})
= \argminn{(\btheta_{\hat\bz}, \bgamma_{\hat\bz})}
\call(\btheta_{\hat\bz}, \bgamma_{\hat\bz})$.

In \cite{hu-compressed} a similar optimization problem involving only the
frequencies is solved by Newton's method. Directly applying that approach in
our case leads to high computational complexity. Methods based on gradient
descent have also been proposed \cite{dublanchet-doa}, but we have observed
that using this approach leads to slow converge. As we are concerned with computational speed
in this paper, we instead use the limited memory
Broyden-Fletcher-Goldfarb-Shanno (L-BFGS) algorithm \cite{nocedal-wright}. This
algorithm only requires evaluation of the objective function and its
gradient. In the following we demonstrate how these evaluations can be
performed with low complexity. At the same time the per-iteration of L-BFGS is
linear in $\hat K$, namely $\calo(J\hat K)$, where $J$ is the number of saved
updates used in L-BFGS. In our implementation we use $J=10$. We have observed
that L-BFGS converges in a small number of iterations.

The L-BFGS algorithm requires an initial estimate of the Hessian of
$\call(\btheta_{\hat\bz}, \bgamma_{\hat\bz})$, which
is subsequently updated in each iteration of the algorithm. Every update of the
activation variable $\hat\bz$ results in a change in the dimension of the
Hessian (the number of variables in $\call(\btheta_{\hat\bz},
\bgamma_{\hat\bz})$ changes). This means that the implicit estimate of the
Hessian in the L-BFGS algorithm is reinitialized rather frequently in our
estimation scheme. As a result, the degree of accuracy of the initialization
of the Hessian has a significant impact on the convergence speed of the
algorithm.  We therefore propose to initialize L-BFGS with a diagonal
approximation of the Hessian. As shown below, the diagonal entries of the
Hessian can be obtained with low computational complexity.

The initial estimate of the Hessian must be positive definite.
This is only achieved when all diagonal entries are positive. Those entries of
the diagonal Hessian which are negative are therefore replaced with the
following values: For entries corresponding to frequency variables we use
$(50N)^2$ as the diagonal Hessian and for the entries corresponding to the
variance of the $k$th component we use $[\hat\bgamma_{\hat\bz}]_k^{-2}$. These
heuristic values have been determined by considering a diagonally scaled
version of the optimization problem (see \cite[Sec.  1.3]{bertsekas-nonlinear}).


Here follows the required first- and second-order partial derivatives of
$\call(\btheta_{\hat\bz}, \bgamma_{\hat\bz})$ evaluated at the current
estimates $(\hat\btheta_{\hat\bz}, \hat\bgamma_{\hat\bz})$ (see
\cite{hansen-sam} for some hints on how these are obtained):
\begin{align*}
	\frac{\partial\call}{\partial [\btheta_{\hat\bz}]_k}
		&= 2[\hat\bgamma_{\hat\bz}]_k
		\im\!\left\{t_k - q_k^* r_k\right\}
		\shownumber \label{derivative} \\
	\frac{\partial\call}{\partial [\bgamma_{\hat\bz}]_k}
		&= s_k - |q_k|^2
		\shownumber \\
	\frac{\partial^2\call}{\partial [\btheta_{\hat\bz}]_k^2}
		&= 2[\hat\bgamma_{\hat\bz}]_k
			\mathrm{Re}\big\{
				x_k - v_k + [\hat\bgamma_{\hat\bz}]_k(t_k^2 - x_ks_k) \\
			&\hspace{10mm}
				+ [\hat\bgamma_{\hat\bz}]_k
					( x_k|q_k|^2 + s_k|r_k|^2 - 2t_kr_kq_k^* ) \\
			&\hspace{33mm}\hfill
				+ (u_kq_k^* - |r_k|^2) \big\}
			\shownumber \label{derivative2} \\
	\frac{\partial^2\call}{\partial [\bgamma_{\hat\bz}]_k^2}
		&=  2s_k|q_k|^2 - s_k^2,
		\shownumber
\end{align*}
where we have defined vectors
\begin{align}
	\bq &\triangleq
		\bPsi\h(\hat\btheta_{\hat\bz}) \bPhi\h \hat\bC\ii \by
		\label{q} \\
	\br &\triangleq
		\bPsi\h(\hat\btheta_{\hat\bz}) \bD \bPhi\h \hat\bC\ii \by
		\label{r} \\
	\bs &\triangleq
		\diag\!\left( \bPsi\h(\hat\btheta_{\hat\bz}) \bPhi\h \hat\bC\ii
		\bPhi \bPsi(\hat\btheta_{\hat\bz}) \right)
		\label{s} \\
	\bt &\triangleq
		\diag\!\left( \bPsi\h(\hat\btheta_{\hat\bz}) \bD \bPhi\h \hat\bC\ii
		\bPhi \bPsi(\hat\btheta_{\hat\bz}) \right)
		\label{t} \\
	\bu &\triangleq
		\bPsi\h(\hat\btheta_{\hat\bz}) \bD^2 \bPhi\h \hat\bC\ii \by
		\label{u} \\
	\bv &\triangleq
		\diag\!\left( \bPsi\h(\hat\btheta_{\hat\bz}) \bD^2 \bPhi\h \hat\bC\ii
		\bPhi \bPsi(\hat\btheta_{\hat\bz}) \right)
		\label{v} \\
	\bx &\triangleq
		\diag\!\left( \bPsi\h(\hat\btheta_{\hat\bz}) \bD \bPhi\h \hat\bC\ii
		\bPhi \bD \bPsi(\hat\btheta_{\hat\bz}) \right).
		\label{x}
\end{align}
The notation $\diag\!(\cdot)$ denotes a vector composed of the diagonal entries of
the (matrix) argument.
The matrix $\hat\bC$ is that in \eqref{py} evaluated at
$\hat\btheta_{\hat\bz}$, $\hat\bgamma_{\hat\bz}$ and $\hat\beta$. We have
defined the diagonal matrix $\bD\triangleq\diag\!\left(
[0,2\pi,4\pi,\ldots,(N-1)2\pi]\T \right)$. In Secs.  \ref{sec:superfast} and
\ref{sec:semifast} we discuss how the vectors \eqref{q}--\eqref{x} can be
calculated with low computational complexity.

\subsubsection{Estimation of activation probability}
With all other variables fixed, the objective \eqref{obj} is a convex function
of $\zeta\in[0,1/2]$
$\left(\frac{\partial^2\call}{\partial\zeta^2}>0\right)$. The global minimizer is then found by
differentiating and setting equal to zero. Considering the constraints on
$\zeta$, we update it as
\begin{align}
	\hat\zeta = \min\!\left(
		\frac{1}{2},
		\frac{||\hat\bz||_0}{K_|max|}
	\right).
	\label{zetaUpdate}
\end{align}

\subsubsection{Estimation of noise variance}
Even when keeping all remaining variables fixed at their current estimate, the
globally minimizing noise variance $\beta$ in \eqref{obj} cannot be found in
closed form. An obvious alternative approach would be to incorporate the estimation of
$\beta$ into L-BFGS together with the estimation of $\btheta_{\hat\bz}$ and
$\bgamma_{\hat\bz}$. However, we have observed this approach to exhibit slow
convergence because the objective function can be rather ``flat'' in the
variable $\beta$ (the gradient is small far away from any stationary point).

In sparse Bayesian learning \cite{wipf-sbl} a similar estimation problem is
solved successfully via the expectation-minimization (EM) algorithm. To
use EM, we need to reintroduce $\balpha$ into the estimation problem. In order
to show how EM is integrated into our coordinate-block descent method and that
the update of $\hat\beta$ is guaranteed not to increase \eqref{obj}, it
is the easiest to directly use the upper bound associated with EM (see
\cite{dellaert-em} for a derivation of EM which takes a similar approach).

The updated estimate of $\beta$ is the minimizer of an upper bound
on the objective function \eqref{obj}. To obtain the upper bound we write the
terms of the objective function which depend on $\beta$, with all
other variables kept fixed at their current estimates:
\begin{align*}
	\call(\beta)
	&= -\ln p(\by|\hat\bz, \hat\btheta; \beta,
		\hat\bgamma) + \const \\
	&= -\ln \int f(\balpha_{\hat\bz})
		\frac{p(\by,\balpha_{\hat\bz}| \hat\bz, \hat\btheta;
			\beta, \hat\bgamma)}{f(\balpha_{\hat\bz})}
		\,\dd\balpha_{\hat\bz} + \const \\
	&\le - \int f(\balpha_{\hat\bz}) 
		\ln\frac{p(\by,\balpha_{\hat\bz}| \hat\bz, \hat\btheta;
			\beta, \hat\bgamma)}{f(\balpha_{\hat\bz})}
		\,\dd\balpha_{\hat\bz} + \const,
		\shownumber\label{bound}
\end{align*}
where $f(\balpha_{\hat\bz})\ge0$ is a function which fulfills $\int
f(\balpha_{\hat\bz})\,\dd\balpha_{\hat\bz}=1$. The inequality follows from
Jensen's inequality.

Following EM, we select
$f(\balpha_{\hat\bz}) =
p(\balpha_{\hat\bz}| \by, \hat\bz, \hat\btheta; \hat\beta^{i-1}, \hat\bgamma)$,
where $\hat\beta^{i-1}$ denotes the previous noise variance
estimate.
Denote the upper bound on the right-hand side of \eqref{bound} by
$Q(\beta; \hat\beta^{i-1})$ and insert
$f(\balpha_{\hat\bz})$ to get
\begin{multline}
	Q(\beta; \hat\beta^{i-1})
	= M\ln\beta + \beta\ii
	\tr\big(\hat\bSigma\bA\h(\hat\btheta_{\hat\bz})\bA(\hat\btheta_{\hat\bz})\big) \\
	+ \beta\ii ||\by - \bA(\hat\btheta_{\hat\bz}) \hat\bmu||^2
	+ \const,
	\label{bound_explicit}
\end{multline}
where we have used \eqref{palpha} to evaluate expectations involving
$\balpha_{\hat\bz}$ and $\hat\bmu$ and $\hat\bSigma$ are calculated from
\eqref{bmu}--\eqref{bSigma} based on $\hat\beta^{i-1}$. It is easy to
show that the upper bound has a unique minimizer, which is used as the updated
estimate of the noise variance:
\begin{align}
	\hat\beta^{i} =
	\max\!\left(\varepsilon_\beta,
	\frac{
		\tr\big(\hat\bSigma\bA\h(\hat\btheta_{\hat\bz})\bA(\hat\btheta_{\hat\bz})\big)
	+ ||\by - \bA(\hat\btheta_{\hat\bz}) \hat\bmu||^2 }{M}
		\right).
	\label{betaUpdate}
\end{align}
To allow low-complexity calculation of $\hat\beta^i$ we use
Woodbury's matrix inversion identity to show that
$\hat\bmu=\hat\bgamma_{\hat\bz}\odot\bq$ and
$\tr\big(\hat\bSigma\bA\h(\hat\btheta_{\hat\bz})\bA(\hat\btheta_{\hat\bz})\big) =
\sum_{k=1}^{\hat K} (\hat\beta^{i-1} s_k [\hat\bgamma_{\hat\bz}]_k)$.

The update \eqref{betaUpdate} could be applied repeatedly since an improved
upper bound is used each time. Since we have not observed any advantages by
doing so, we simply perform the update \eqref{betaUpdate} once for each pass
in the block-coordinate descent algorithm. We also note that even though EM is
known to be prone to slow convergence speed, we have observed empirically that
the estimate of $\beta$ converges fast, typically within 10 iterations.

It can easily be shown that with the chosen $f(\balpha_{\hat\bz})$, the
inequality in \eqref{bound} holds with equality at $\beta=\hat\beta^{i-1}$.
It then follows that the new estimate of $\beta$ does not increase the value of
the objective function (see the proof of Lemma \ref{lem:beta} in Appendix
\ref{app:convergence}).

\subsubsection{Deactivation of Components}
\label{sec:deactivate}
We now describe the activation and deactivation of components, which is
performed by the single most likely replacement (SMLR) detector
\cite{kormylo-ml}. SMLR has previously been demonstrated to perform well
for LSE \cite{hansen-sam, badiu-valse,champagnat-unsupervised,
dublanchet-doa}.

First we write the terms of \eqref{obj} which depend on the
variables pertaining to the $k$th component and fix all other variables at
their current estimate. Based on Woodbury's matrix inversion identity and the
determinant lemma we get (see \cite{tipping-fastmarginal} for details)
\begin{multline*}
	\call(z_k, \theta_k, \gamma_k)
	= z_k
	\Bigg(
		- \frac{|q_{\sim k}(\theta_k)|^2}
			{\gamma_k\ii + s_{\sim k}(\theta_k)}
		\\
		+ \ln\!\left(
			\left( 1 + \gamma_ks_{\sim k}(\theta_k) \right)
			\frac{1-\hat\zeta}{\hat\zeta}
		\right)
	\Bigg)
	+ \const, \shownumber \label{singlecomp}
\end{multline*}
with
\begin{align*}
	q_{\sim k}(\theta_k) &\triangleq \bpsi\h(\theta_k)\bPhi\h \hat\bC_{\sim k}\ii \by
	\shownumber \label{qTilde} \\
	s_{\sim k}(\theta_k) &\triangleq \bpsi\h(\theta_k)\bPhi\h \hat\bC_{\sim k}\ii
	\bPhi\bpsi(\theta_k),
\end{align*}
where $\hat\bC_{\sim k} \triangleq \hat\beta\bI + \bA(\hat\btheta_{\hat\bz_{\sim k}})
\hat\bGamma_{\hat\bz_{\sim k}} \bA\h(\hat\btheta_{\hat\bz_{\sim k}})$ and
$\hat\bz_{\sim k}$ is equal to $\hat\bz$ with the $k$th entry forced to zero. The
matrix $\hat\bC_{\sim k}$ is thus the marginal covariance matrix of the
observation vector with the $k$th component deactivated.

To evaluate if an active component should be deactivated, we test if the
objective \call is increased by doing so, i.e., we test if $\call(z_k=0,
\hat\theta_k, \hat\gamma_k) < \call(z_k=1, \hat\theta_k, \hat\gamma_k)$. This
gives the deactivation criterion for the $k$th component:
\begin{align}
	\frac{|q_{\sim k}(\hat\theta_k)|^2}{\hat\gamma_k\ii +
	s_{\sim k}(\hat\theta_k)}
	- \ln\!\left(1+\hat\gamma_k s_{\sim k}(\hat\theta_k)\right)
	< \ln\!\left(\frac{1-\hat\zeta}{\hat\zeta}\right).
	\label{deactivate}
\end{align}
This criterion is evaluated for currently active components, i.e., for $k$
which has corresponding $\hat z_k=1$.

For computational convenience we note that we can obtain $q_{\sim
k}(\hat\theta_k)$ and $s_{\sim k}(\hat\theta_k)$ from $\bq$ and
$\bs$ with low complexity. First, write $\hat\bC_{\sim k} = \hat\bC -
\hat\gamma_k\bPhi\bpsi(\hat\theta_k)\bpsi\h(\hat\theta_k)\bPhi\h$ and use
Woodbury's identity to obtain
\begin{align*}
	q_{\sim k}(\hat\theta_k) &=
		\frac{q_i}{1 - \hat\gamma_k s_i} \\
	s_{\sim k}(\hat\theta_k) &=
		\frac{s_i}{1 - \hat\gamma_k s_i},
\end{align*}
where $q_i$ and $s_i$ are the $i$th entries of \eqref{q} and \eqref{s} with
$i$ denoting the index for which $[\hat\btheta_{\hat\bz}]_i =
\hat\theta_k$.

\subsubsection{Component Activation}
\label{sec:activate}
We now describe a method to decide if a deactivated component should be
activated. This also involves estimating the frequency and variance of this
component, because no meaningful such estimates are available before the
component is activated. Any of the deactivated components are equally good
candidates for activation. In the following $k$ refers to an arbitrary value for
which $\hat z_k=0$. If no such $k$ exists all components are already activated
and the activation step is not carried out.

Our method is again based on the expression \eqref{singlecomp}. Inspired by
\cite{badiu-valse}, let $\bar\gamma$ denote the average of the entries in
$\hat\bgamma_{\hat\bz}$. Define the change in the objective obtained from
setting $\hat z_k=1$, $\hat\theta_k=\theta_k$, $\hat\gamma_k=\bar\gamma$:
\begin{align*}
	\Delta\call(\theta_k) &= \call(1,\theta_k,\bar\gamma) -
	\call(0,\theta_k,\bar\gamma)
	\shownumber\label{delta_L} \\
	&= \ln\!\left(
		\left( 1 + \bar\gamma s_{\sim k}(\theta_k) \right)
		\frac{1-\hat\zeta}{\hat\zeta}
	\right)
	- \frac{|q_{\sim k}(\theta_k)|^2}
		{\bar\gamma\ii + s_{\sim k}(\theta_k)}
\end{align*}
Note that the last term in \eqref{delta_L} does not depend on $\theta_k$ or $\bar\gamma$. Then the
frequency is found by maximizing the decrease in the objective, i.e.,
\begin{align}
	\hat\theta_k &= \argminn{\theta_k\in\calg} \Delta\call(\theta_k),
		\label{activate_theta_argmin}
\end{align}
where $\calg$ is a grid of $L$ equispaced values, i.e.,
$\calg\triangleq\{0,1/L,\ldots,1-1/L\}$.
The restriction of the estimated frequencies to a grid does not mean that the
final frequency estimates lie on a grid, because they are refined
to be in $[0,1)$ in subsequent updates of the frequency vector.
For this reason, the choice of $L$ does not have any impact on the estimation
accuracy, provided that it is sufficiently large.%
\footnote{
A numerical investigation (not reported here) shows that the algorithm is
invariant to the choice of $L$, provided that $L\ge2N$. In our implementation
we use $L$ equal to $8N$ rounded to the nearest power of 2.
}
In Sec. \ref{sec:superfast} and \ref{sec:semifast} we show
how $q_{\sim k}(\theta_k)$ and $s_{\sim k}(\theta_k)$ can be evaluated with low
complexity for all $\theta_k\in\calg$, such that the minimization can be
performed by means of an exhaustive search over $\calg$.

The activation procedure continues only if a decrease in the objective can be
obtained by activating a component at $\hat\theta_k$, i.e., if
$\Delta\call(\hat\theta_k)<-\varepsilon_\call$. The inclusion of the constant
$\varepsilon_\call>0$ is purely technical, as it simplifies our convergence
analysis. It can be chosen arbitrarily small and we select it as machine
precision in our implementation.

After estimating the frequency, the component variance is selected as
$\hat\gamma_k = \argminn{\gamma_k} \Delta\call(1,\hat\theta_k, \gamma_k)$.
Using an approach similar to \cite{tipping-fastmarginal}, this minimizer can be
shown to be
\begin{align*}
	\hat\gamma_k &=
	\begin{cases}
		\frac{|q_{\sim k}(\theta_k)|^2 - s_{\sim k}(\theta_k)}{s_{\sim k}^2(\theta_k)}
			& \textrm{if }\frac{|q_{\sim k}(\theta_k)|^2}{s_{\sim k}(\theta_k)} > 1, \\
		0	& \textrm{otherwise}.
	\end{cases}
	\shownumber\label{activate_gamma}
\end{align*}
The component is only activated if%
\footnote{When $\hat\gamma_k=0$ the $k$th component is effectively deactivated
because the corresponding coefficient $\alpha_k$ has a zero-mean prior with
zero variance, see \eqref{alphaPrior}. The effective deactivation is also seen in
the definition of \bC in \eqref{py} and it further manifests itself as
$\hat\mu_k=0$ in \eqref{bmu}.}
$\hat\gamma_k>0$.

It is instructive to explore the activation criterion
$\Delta\call(\hat\theta_k) < - \varepsilon_\call$ in detail. Since
$\varepsilon_\call$ is machine precision, we ignore it ($\varepsilon_\call=0$)
for simplicity. The activation criterion can be rewritten to the form
\begin{align}
	\frac{|q_{\sim k}(\hat\theta_k)|^2}
		{s_{\sim k}(\hat\theta_k)}
	>
	\left( 1 + \frac{1}{\bar\gamma s_{\sim k}(\hat\theta_k)} \right)
	\ln\!\left(
		\left( 1 + \bar\gamma s_{\sim k}(\hat\theta_k) \right)
		\frac{1-\hat\zeta}{\hat\zeta}
	\right).
	\label{activate_derived}
\end{align}
Denote the left-hand side of \eqref{activate_derived} as $\kappa_k$. This
quantity can be interpreted as the signal-to-noise ratio of the $k$th component
\cite{shutin-artifact,shutin-fvsbl}. If the sparse Bayesian learning (SBL) model is used for
sparsity promotion, an activation criterion of the from $\kappa_k>1$ is
obtained \cite{shutin-artifact,shutin-fvsbl}. Algorithms using the activation criterion
$\kappa_k>1$ are known to be prone to the activation of ``artefact'' components
with very small $\hat\gamma_k$ and $\hat\alpha_k$ at what seems to be arbitrary
frequencies $\hat\theta_k$.  The right-hand side of \eqref{activate_derived} is
always larger than one and this helps reduce the number of artefacts which
are activated, as demonstrated in \cite{badiu-valse}. This favorable
phenomenon is caused by the use of the average $\bar\gamma$ in the definition
of $\Delta\call(\theta_k)$ (as opposed to inserting $\hat\gamma_k$ from
\eqref{activate_gamma}, which resembles the SBL approach).

Even still, we have observed the activation of a few artefact components in our
numerical investigations. We therefore follow the same idea as
\cite{shutin-artifact,shutin-fvsbl} and heuristically adjust the criterion
\eqref{activate_derived} to obtain
{\small
\begin{align}
	\frac{|q_{\sim k}(\theta_k)|^2}
		{s_{\sim k}(\theta_k)}
	>
	\left( 1 + \frac{1}{\bar\gamma s_{\sim k}(\theta_k)} \right)
	\ln\!\left(
		\left( 1 + \bar\gamma s_{\sim k}(\theta_k) \right)
		\frac{1-\hat\zeta}{\hat\zeta}
	\right) + \tau,
	\label{activate_criterion}
\end{align}
}%
where $\tau\ge0$ is some adjustment of the threshold. Specifically we select
$\tau=5$, cf. the numerical study in Sec.  \ref{sec:threshold}. Our numerical
experiments show that this simple approach is very effective at avoiding the
inclusion of small spurious components. Since the heuristic criterion
\eqref{activate_criterion} is stricter than the criterion
$\Delta\call(\hat\theta_k)<-\varepsilon_\call$, it is guaranteed that the
activation of a component decreases the objective function.

\subsection{Outline of the Algorithm and Implementation Details}
\label{sec:outline}
The algorithm proceeds by repeating the following steps until convergence:
\begin{enumerate}
	\item \label{step:activate}
		Check if any components can be activated via the procedure described
		in Sec. \ref{sec:activate}.
	\item \label{step:zeta}
		Re-estimate the activation probability $\zeta$ via \eqref{zetaUpdate}.
	\item \label{step:beta}
		Re-estimate the noise variance via \eqref{betaUpdate}.
	\item \label{step:repeat}
		Repeat:
		\begin{enumerate}
			\item[\ref{step:repeat}a)] \label{step:theta_gamma}
				Perform a single L-BFGS update of the estimated vectors of
				active component frequencies $\btheta_{\hat\bz}$ and variances
				$\bgamma_{\hat\bz}$, as described in Sec. \ref{sec:thetaUpdate}.
			\item[\ref{step:repeat}b)]\label{step:deact}
				Check if any components can be deactivated via \eqref{deactivate}.
			\end{enumerate}
\end{enumerate}
The algorithm terminates when the change in the objective \eqref{obj} between
two consecutive iterations is less than $M10^{-7}$.

In step \ref{step:activate}) and \ref{step:deact}b) the check for component
(de)activation is repeated until no more components can be (de)activated.
The updates in step \ref{step:repeat}) are iterated until either the
approximated squared Newton decrement of the L-BFGS method is below $M10^{-8}$
or at most 5 times.

The observant reader will have noticed that the minimization over
$(\btheta_{\hat\bz},\bgamma_{\hat\bz})$ must be constrained to $\gamma_k\ge0$
for all $k$. It turns out that this constraint can be handled in a simple
manner: Notice that the deactivation criterion \eqref{deactivate} is always
fulfilled for $\hat\gamma_k$ sufficiently small. The constraint is therefore
never active at the solution. We therefore simply need to
restrict the line-search performed in L-BFGS such that the no entry in
$\hat\bgamma_{\hat\bz}$ ever becomes negative. If any $\hat\gamma_k$ approaches
(or becomes equal to) zero, it is deactivated in step \ref{step:deact}b). Note
that this approach resembles that of L-BFGS for box constraints
\cite{byrd-lbfgs-b}, except that the deactivation of variables for which the
constraint is active happens automatically in our algorithm.

The algorithm is initialized with all components in the deactivated stage (i.e.
$\hat\bz=\mathbf{0}$). The initial values of the entries in $\hat\btheta$ and
$\hat\bgamma$ do not matter, since they are assigned when their corresponding
component is activated (see Sec.  \ref{sec:activate}). The noise variance is
initialized to $\hat\beta=0.01||\by||^2/M$ ($1\,\%$ of the energy in \by is
assumed to be noise). The activation probability is initialized to
$\hat\zeta=0.2$.

In Appendix \ref{app:convergence} we discuss in detail the convergence
properties of our algorithm. The findings are summarized here. We show that our
algorithm terminates in finite time and that the estimates of
$\bz$, $\zeta$ and $\beta$ are guaranteed to converge. We denote the limit
points as $\bar\bz$, $\bar\zeta$ and $\bar\beta$. When these estimates have
converged, our algorithm reduces to a pure L-BFGS scheme which estimates
$(\btheta_{\bar\bz}, \bgamma_{\bar\bz})$.
Due to the non-convexity of the
objective function, we cannot guarantee convergence of L-BFGS (see
\cite{mascarenhas-bfgs}). Despite of this, we have never observed
non-convergence of our algorithm. In our experiments it always
converged to a local minimum of the objective function. We therefore rely on
the vast amount of experimental validation of the convergence of L-BFGS and
assume convergence to a stationary point. In particular, we have the following
theorem.
\begin{theorem}
	\label{thm:convergence}
	Assume that L-BFGS in step 4a) converges to a stationary
	point of $(\btheta_{\bar\bz}, \bgamma_{\bar\bz})\mapsto
	\call(\bar\bz, \bar\zeta, \bar\beta, \btheta_{\bar\bz},
	\bgamma_{\bar\bz})$. Then the sequence of estimates obtained by our
	algorithm converges.
	Further, the limit point is a stationary point of
	$(\zeta,\beta,\btheta,\bgamma)\mapsto
	\call(\bar\bz, \zeta, \beta, \btheta, \bgamma)$,
	in the sense that the Karush-Kuhn-Tucker necessary conditions for a minimum
	are fulfilled.
\end{theorem}
\begin{IEEEproof}
	See Appendix \ref{app:convergence}.
\end{IEEEproof}


\subsection{Initial Activation of Components}
When the number of sinusoids $K$ in the observed signal \eqref{model} is high,
the algorithm spends significant computational effort activating components (step
\ref{step:activate}). This is because each time a component is activated, the
values $q_{\sim k}(\theta_k)$ and $s_{\sim k}(\theta_k)$ must be evaluated for
all $\theta_k\in\calg$ to calculate \eqref{activate_theta_argmin}.
To alleviate the computational effort of building the initial set of
active components, we propose to let the first few iterations use an approximate
scheme for activating components in place of step \ref{step:activate}). The approximate activation scheme proceeds as follows:
\begin{enumerate}
	\item Calculate $q_{\sim k}(\theta)$ and $s_{\sim k}(\theta)$ for all
		$\theta\in\calg$, where $k$ is the index of a deactivated component.
	\item Evaluate $\Delta\call(\theta)$ \eqref{delta_L} for all
		$\theta\in\calg$.
	\item Find the local minimizers of $\Delta\call(\theta)$, i.e., find the
		values of $\theta$ for which $\Delta\call(\theta)\le
		\Delta\call(\theta')$ with $\theta'$ being any of the two
		neighbouring grid-points of $\theta$. The local minimizers are
		candidate frequencies.
	\item Activate a component at those candidate frequencies for which
		the following criteria are fulfilled:
		\begin{itemize}
			\item The component activation criterion \eqref{activate_derived}
				is fulfilled.
			\item The component variance \eqref{activate_gamma} is non-zero.
			\item The decrease in the objective obeys
				$\Delta\call(\theta)\le\Delta\call_|min|/5$,
				where $\Delta\call_|min|$ is the largest decrease obtained from
				activating a component at another candidate frequency (in the
				current iteration).
			\item All other currently active components have frequency
				estimates located at least%
				\footnote{For the distance measure we use the wrap-around
				distance on $[0,1)$ defined as
				$d(x,y)\triangleq\min(|x-y|, 1-|x-y|)$ for $x,y\in[0,1)$.\label{distance}}
				$0.05N\ii$ apart from the candidate frequency.
		\end{itemize}
\end{enumerate}
The above method is a heuristic scheme, which quickly builds a set of activated
components. Typically this set is close to the final result and only a few (in
our setup less than 15 in most cases) iterations are need before convergence.

\section{Superfast Method (Complete Observations)}
\label{sec:superfast}
The algorithm presented above has rather large computational complexity, in
particular due to the inversion of \bC and the calculation of $q_{\sim
k}(\theta)$ $s_{\sim k}(\theta)$ for all $\theta\in\calg$. In this section we
discuss how all updates of the algorithm can be evaluated with low
computational complexity by exploiting the inherent structure of the problem.
In particular we discuss how to evaluate 
$\ln|\bC|$, $\by\h\bC\ii\by$, $\bq$, $\br$, $\bs$, $\bt$, $\bu$, $\bv$, $\bx$
and $q_{\sim k}(\theta)$, $s_{\sim k}(\theta)$ for all $\theta\in\calg$.

The method presented here is only applicable when the complete observation
vector is available, i.e., when $\bPhi=\bI$, $M=N$ and
$\bA(\btheta)=\bPsi(\btheta)$. In this case the observation vector \by is a
wide-sense stationary process and its covariance matrix \bC is Hermitian
Toeplitz. Low-complexity algorithms for inverting such matrices are available
in the literature.
We also rely on fast Fourier transform (FFT) techniques.

Our approach is based on the Gohberg-Semencul formula
\cite{gohberg-convolution, ammar-generalized}, which states that the inverse of
the Hermitian Toeplitz matrix \bC can be decomposed as
\begin{align}
	\bC\ii = \delta_{N-1}\ii \left( \bT_1\h\bT_1 - \bT_0\bT_0\h
	\right),
	\label{gohberg-semencul}
\end{align}
where the entries of $\bT_0$ and $\bT_1$ are
\begin{align*}
	[\bT_0]_{i,k} &= \rho_{i-k-1}, \\
	[\bT_1]_{i,k} &= \rho_{N-1+i-k}
\end{align*}
for $i,k=1,\ldots,N$. Note that $\rho_i=0$ for $i<0$ and $i>N-1$; thus $\bT_0$ is
strictly lower triangular and $\bT_1$ is unit upper triangular
($\rho_{N-1}=1$). The values $\delta_i$ and $\rho_i$ for $i=0,\ldots,N-1$ can
be computed with a generalized Schur algorithm in time $\calo(N\log^2N)$
\cite{ammar-generalized}. Alternatively, the Levinson-Durbin algorithm can also
be used to obtain the decomposition in time $\calo(N^2)$. The latter algorithm
is significantly simpler to implement and is faster for small $N$. In
\cite{ammar-numerical} it is concluded that the Levinson-Durbin algorithm
requires fewer total operations than the generalized Schur algorithm for
$N\le256$.

\subsection{Evaluating \texorpdfstring{$\by\h\bC\ii\by$ and $\ln|\bC|$}{yHCinvy
and ln C}}
To calculate the value of the objective function \eqref{obj} we need to find
$\by\h\bC\ii\by$ and $\ln|\bC|$. Inspecting \eqref{gohberg-semencul} it is
clear that matrix-vector products involving $\bT_0$ and $\bT_1$ are
convolutions. These can be implemented using FFT techniques. The product
$\by\h\bC\ii\by$ can thus be calculated in $\calo(N\log N)$ time when
$\{\rho_i\}$ and $\delta_{N-1}$ are known.

The matrix \bC is Hermitian positive definite and can therefore be factorized
uniquely as
\begin{align}
	\bC = \bL \bB \bL\h,
\end{align}
with \bL being unit lower triangular. The diagonal matrix \bB is computed with the
generalized Schur algorithm. Its diagonal entries are given by $\delta_i$
for $i=0,\ldots,N-1$ \cite{ammar-generalized}. Since the determinant of a
triangular matrix is the product of its diagonal entries, we have
\begin{align}
	\ln|\bC| = \sum_{i=0}^{N-1} \ln\delta_i.
\end{align}
It follows that once the generalized Schur algorithm has been executed, the
objective function \eqref{obj} can easily be found.

\subsection{Evaluating \texorpdfstring{$\bq$, $\br$ and
$\bu$}{q, r and u}}
Note that $\bC\ii\by$ can be evaluated with FFT techniques using
\eqref{gohberg-semencul}. We recognize that matrix-vector products involving
$\bPsi\h(\hat\btheta_{\hat\bz})$ are Fourier transforms evaluated off the
equispaced grid. Such products are approximated to a very high precision in
time $\calo(N\log N)$ using the non-uniform fast Fourier transform%
\footnote{The NUFFT calculates the Fourier transform at arbitrary points (not
lying on an equispaced grid) by interpolation combined with an FFT.
It is an approximation, which can be made arbitrarily accurate by
including more points in the interpolation. The NUFFT achieves a time
complexity of $\calo(N\log N + K)$, where $K$ is the number of off-the-grid
frequency points at which it is evaluated. For $K\le N$ this complexity is
equal to that of the FFT, but the constant hidden in the big-O
notation is much higher for the NUFFT. We have found that for $N\ge 512$
significant speedups can be achieved by using the NUFFT over a direct
computation of $\bA(\hat\btheta_{\hat\bz})$ and evaluation of the matrix-vector
products involving this matrix. In particular the speedup arises from the fact
that $\bA(\hat\btheta_{\hat\bz})$ no longer needs to be formed.}
(NUFFT) \cite{greengard-nufft, lee-nufft}.
Then $\bq$, $\br$ and $\bu$ are easily found in time $\calo(N\log N)$ (assuming the
decomposition \eqref{gohberg-semencul} has already been calculated).

\subsection{Evaluating \texorpdfstring{$\bs$, $\bt$, $\bv$ and
$\bx$}{s, t, v and x}}
Turning our attention to $\bs$, we follow
\cite{musicus-fastmlm} and note that (recall that we assume $\bPhi=\bI$)
\begin{align*}
	s_k &=
	[\bPsi\h(\hat\btheta_{\hat\bz}) \bC\ii \bPsi(\hat\btheta_{\hat\bz})]_{k,k} \\
	&= \sum_{i=-(N-1)}^{N-1} \omega_s(i)
	\exp\!\left(j2\pi i [\hat\btheta_{\hat\bz}]_k \right)
	\shownumber \label{sFourier}
\end{align*}
for $k=1,\ldots,\hat K$ where $\hat K$ is the number of entries in
$\hat\btheta_{\hat\bz}$. The function $\omega_s(i)$ gives the sum over the
$i$th diagonal, i.e.,
\begin{align}
	\omega_s(i) = \sum_{q=\max(0,-i)}^{\min(N-1-i,N-1)} [\bC\ii]_{q+1,q+i+1}.
	\label{omega_s}
\end{align}
It is obvious that \eqref{sFourier} can be calculated for all $k=1,\ldots,\hat K$
via a NUFFT when the values $\omega_s(i)$ are available.

To evaluate $\bt$, $\bv$ and $\bx$ we follow a similar approach and note that
the entries of these vectors can be written as \eqref{sFourier} with
$\omega_s(i)$ replaced by
\begin{align}
	\omega_t(i) &= \sum_{q=\max(0,-i)}^{\min(N-1-i,N-1)} [\bD\bC\ii]_{q+1,q+i+1}
	\label{omega_t} \\
	\omega_v(i) &= \sum_{q=\max(0,-i)}^{\min(N-1-i,N-1)} [\bD^2\bC\ii]_{q+1,q+i+1}
	\label{omega_v} \\
	\omega_x(i) &= \sum_{q=\max(0,-i)}^{\min(N-1-i,N-1)}
	[\bD\bC\ii\bD]_{q+1,q+i+1},
	\label{omega_x}
\end{align}
respectively. In Appendix \ref{app:superfast} we demonstrate how
$\{\omega_s(i)\}$, $\{\omega_t(i)\}$, $\{\omega_v(i)\}$ and $\{\omega_x(i)\}$
can be obtained through length-$2N$ FFTs using the decomposition
\eqref{gohberg-semencul}.

\subsection{Evaluating \texorpdfstring{$q_{\sim k}(\theta)$ and
$s_{\sim k}(\theta)$ for all $\theta\in\calg$}{q and s at all gridpoints}}
To calculate the frequency of the component processed in the activation stage,
$q_{\sim k}(\theta)$ and $s_{\sim k}(\theta)$ must be evaluated for all
$\theta\in\calg$, where $\calg$ is a grid of $L$ equispaced points. Defining
the vector of gridded frequencies $\btheta^\calg\triangleq[0,1/L,\ldots,
(L-1)/L]\T$, we need to find
\begin{align*}
	\bq^\calg &\triangleq \bPsi\h(\btheta^\calg) \bC\ii \by, \\
	\bs^\calg &\triangleq \diag\!\left( \bPsi\h(\btheta^\calg) \bC\ii
	\bPsi(\btheta^\calg) \right).
\end{align*}
We have used the fact that in the beginning of the activation step the $k$th
component is deactivated and thus $\bC_{\sim k} = \bC$.

Since $\calg$ is an equispaced grid, products with $\bPsi\h(\btheta^\calg)$ can
be evaluated as a length-$L$ FFT. The vector $\bq^\calg$ is therefore easy to
find. Rewriting $\bs^\calg$ in the form \eqref{sFourier}, it is seen that
$\bs^\calg$ can also be evaluated as a length-$L$ FFT. These computations have
time-complexity $\calo(L\log L)$ (assuming the decomposition
\eqref{gohberg-semencul} has already been calculated).

\subsection{Algorithm Complexity}
In summary, the time complexity of each iteration in the algorithm described in
Sec. \ref{sec:algorithm} is dominated by either the calculation of $\{\rho_i\}$
and $\delta_{N-1}$ with the generalized Schur algorithm or the calculation of
$\bq^\calg$ and $\bs^\calg$ (we assume $\hat K \le M = N\le L$).
With our choice $L=8N$ we have complexity
per iteration of $\calo(N\log^2 N)$.

Also note that all computations involving $\bPsi(\hat\btheta_{\hat\bz})$
are performed using the NUFFT. This matrix therefore does not need to be
stored, so our algorithm only uses a modest amount of memory.

\section{Semifast Method (Incomplete Observations)}
\label{sec:semifast}
The method presented in Sec. \ref{sec:superfast} is not applicable when an
incomplete observation vector is available, i.e., when $\bPhi\ne\bI$. In the
following we introduce a computational method, which can be used when \bPhi is
a subsampling and scaling matrix, i.e., when $\bPhi\in\bbc^{M\times N}$
consists of $M$ rows of a diagonal matrix.%
\footnote{Let $\calm\subseteq\{1,\ldots,N\}$ denote the index set
of the observed entries and 
$I_\calm : \{1,\ldots,M\} \rightarrow \calm$ be an indexing. Then
$\bPhi_{m,{I_\calm(m)}}$, $m=1,\ldots,M$, are the only nonzero elements of
$\bPhi$.}
With this method we can still obtain an
algorithm with reasonable computational complexity per iteration, assuming that
$\hat K$ is relatively small (a $\hat K \times \hat K$ matrix must be
inverted). We coin this algorithm as semifast.
For small $\hat K$ the semifast algorithm is faster than the superfast
algorithm of Sec.  \ref{sec:superfast} and it may therefore be beneficial to
even use it in the complete data case.

The semifast method is based on the following decomposition of $\bC\ii$,
obtained using Woodbury's matrix identity:
\begin{align}
	\bC\ii &= \hat\beta\ii\bI - \hat\beta^{-2}\bA(\hat\btheta_{\hat\bz}) \hat\bSigma
	\bA\h(\hat\btheta_{\hat\bz})
	\label{CfromSigma}
\end{align}
with $\hat\bSigma$ given by \eqref{bSigma}. We can evaluate $\hat\bSigma\ii$ by noting
that
\begin{multline*}
	\left[ \bA\h(\hat\btheta_{\hat\bz})\bA(\hat\btheta_{\hat\bz}) \right]_{i,k}
	= \left[ \bPsi\h(\hat\btheta_{\hat\bz}) \bPhi\h\bPhi
		\bPsi(\hat\btheta_{\hat\bz}) \right]_{i,k} \\
	= \sum_{m=1}^M |\bPhi_{m,I_\calm(m)}|^2
		\exp\!\left( j2\pi(I_\calm(m)-1)(\hat\theta_k-\hat\theta_i) \right),
		\shownumber \label{AhA}
\end{multline*}
which can be evaluated with a NUFFT in time $\calo(N\log N + \hat K^2)$. 
Forming $\hat\bSigma\ii$ is then easy and an inversion%
\footnote{As is customary in numerical linear algebra, we would recommend not
to explicitly evaluate the inverse, but instead use the numerically stabler and
faster approach of calculating the Cholesky decomposition $\hat\bSigma\ii=\bL\bL\h$
(a unique Cholesky decomposition exists because $\hat\bSigma\ii$ is Hermitian
positive definite). We need to evaluate matrix-vector products involving
$\hat\bSigma$ which are easily evaluated from the decomposition by forward-backward
substitution. We can also calculate $|\hat\bSigma\ii|$ directly from the Cholesky decomposition.}
in time $\calo(\hat K^3)$ is needed to obtain $\hat\bSigma$. The approach thus hinges
on $\hat K$ being sufficiently small, such that the inverse (really, the
Cholesky decomposition) can be calculated in reasonable time.

\subsection{Evaluating \texorpdfstring{$\by\h\bC\ii\by$, $\ln|\bC|$,
$\bq$, $\br$ and $\bu$}{yHCinvy, ln C, q, r and u}}
Notice that matrix-vector products involving $\bPsi(\hat\btheta_{\hat\bz})$ and
$\bPsi\h(\hat\btheta_{\hat\bz})$ can be evaluated using a NUFFT. It then
immediately follows that the values $\by\h\bC\ii\by$, $\bq$, $\br$ and $\bu$
can be evaluated using \eqref{CfromSigma} with complexity $\calo(\hat K^2 +
N\log N)$.

To evaluate the objective function \eqref{obj} we need to calculate $\ln|\bC|$.
By invoking the matrix determinant lemma we get
\begin{align}
	\ln|\bC|
	&=
	M\ln\hat\beta 
	+ \smashoperator{\sum_{\{k:\hat z_k=1\}}}\: \ln \hat\gamma_k
	+ \ln|\bSigma\ii|,
\end{align}
which can be evaluated in time $\calo(\hat K)$ once the Cholesky
decomposition of $\bSigma\ii$ is known.

\subsection{Evaluating \texorpdfstring{$\bs$, $\bt$, $\bv$ and
$\bx$}{s, t, v and x}}
As an example, we demonstrate how to evaluate $\bt$. We note that $\bs$, $\bv$
and $\bx$ can easily be obtained using the same approach.
First, insert \eqref{CfromSigma} into \eqref{t} to get
\begin{multline*}
	\bt =
	\hat\beta\ii \diag\left( \bPsi\h(\hat\btheta_{\hat\bz}) \bD \bPhi\h
		\bA(\hat\btheta_{\hat\bz}) \right) \\
	- \hat\beta^{-2} \diag\left( \bPsi\h(\hat\btheta_{\hat\bz}) \bD \bPhi\h
		\bA(\hat\btheta_{\hat\bz}) \hat\bSigma
		\bA\h(\hat\btheta_{\hat\bz})
		\bA(\hat\btheta_{\hat\bz}) \right).
\end{multline*}
Using the same methodology as for computing $\hat\bSigma\ii$, the $\hat K\times
\hat K$ matrices $\bA\h(\hat\btheta_{\hat\bz})\bA(\hat\btheta_{\hat\bz})$ and
$\bPsi\h(\hat\btheta_{\hat\bz}) \bD \bPhi\h \bA(\hat\btheta_{\hat\bz})$
can be obtained in time $\calo(N\log N + \hat K^2)$. Then, $\bt$ is found by
direct evaluation in time $\calo(\hat K^3)$.

\subsection{Evaluating \texorpdfstring{$q_{\sim k}(\theta)$ and
$s_{\sim k}(\theta)$ for all $\theta\in\calg$}{q and s at all gridpoints}}
To calculate the frequency of the component processed in the activation stage
we must evaluate $q_{\sim k}(\theta)$ and $s_{\sim k}(\theta)$ for all
$\theta\in\calg$, where $\calg$ is a grid of $L$ equispaced points.
Using the fact that in the beginning of the activation step the $k$th
component is deactivated and thus $\bC_{\sim k} = \bC$, we obtain the required
quantities by inserting \eqref{CfromSigma} into \eqref{q} and \eqref{s}:
\begin{align*}
	\bq^\calg &=
		\hat\beta\ii \bA\h(\btheta^\calg)
		\left( \by - \bA(\hat\btheta_{\hat\bz}) \hat\bmu
	\right) \\
	\bs^\calg &=
		\hat\beta\ii \diag\!\left( \bA\h(\btheta^\calg)
		\bA(\btheta^\calg) \right) \\
		& \hspace{5mm}
		- \hat\beta^{-2} \diag\!\left( \bA\h(\btheta^\calg)
		\bA(\hat\btheta_{\hat\bz}) \hat\bSigma
		\bA\h(\hat\btheta_{\hat\bz})
		\bA(\btheta^\calg) \right).
\end{align*}
It is clear that $\bq^\calg$ can easily be found using FFT techniques.

To obtain $\bs^\calg$ we first note that the first term is a vector with all
entries equal to $\hat\beta\ii\sum_{m=1}^M |\bPhi_{m,I_\calm(m)}|^2$. The second
term is found by using a NUFFT (see \eqref{AhA}) to form 
$\bA\h(\hat\btheta_{\hat\bz}) \bA(\btheta^\calg)$. Then by using the Cholesky
decomposition of $\hat\bSigma\ii$ the second term can be calculated in time
$\calo(L\hat K^2)$.



\subsection{Algorithm Complexity}
The above computation is dominated by either the calculation of $\bs^\calg$ or
the length-$L$ FFT involved in calculating $\bq^\calg$. Again with $L=8N$ we
have overall complexity per iteration $\calo(N\hat K^2 + N\log N)$.

\section{Multiple Measurement Vectors}
\label{sec:mmv}
The algorithm presented in Sec. \ref{sec:alg} assumes a single measurement
vector (SMV). We now discuss an extension to the case of multiple
measurement vectors (MMV) \cite{cotter-mmv}. This case is of particular
importance in array processing where the number of observation points $M$ is
determined by the number of antennas in the array.%
\footnote{It is worth noting that in array processing the complete data case
corresponds to the very common situation of using a uniform linear array.}
Typically $M$ is small, which thus limits estimation accuracy. On the
other hand it is often easy to obtain multiple observation vectors across
which the entries in $\tilde\btheta$ (the true directions of arrivals) are practically
unchanged. The MMV signal model reads
\begin{align}
	\by^{(g)} = \bA(\tilde\btheta)\tilde\balpha^{(g)} + \bw^{(g)},
\end{align}
where $g=1,\ldots,G$ indexes the observation vectors.

To extend our SMV algorithm to the MMV case we again impose an estimation
model of the form \eqref{estModel} that contains $K_|max|$ components which can
be (de)activated based on variables $z_k$, $k=1,\ldots,K_|max|$. The
likelihood for each of the $G$ observation vectors then reads
\begin{align}
	p(\by^{(g)} | \balpha^{(g)}, \bz, \btheta; \beta) = \CN(\by^{(g)}; \bA(\btheta_\bz)
	\balpha_\bz^{(g)}, \beta\bI).
\end{align}
We impose the same prior as used in the SMV case \eqref{alphaPrior} on each
$\balpha^{(g)}$:
\begin{align}
	p(\balpha^{(g)}; \bgamma) = \prod_{k=1}^{K_|max|} \CN(\alpha_k^{(g)}; 0,
	\gamma_k).
\end{align}
The vectors $\bz$ and $\btheta$ are assigned the same priors as in the SMV
case, i.e., as given by \eqref{zPrior} and \eqref{thetaPrior}. Similarly to the
SMV case, the MMV model has parameters $\bgamma$, $\beta$ and $\zeta$.

The objective to be minimized is the marginal likelihood, which for
the MMV model reads
\begin{align*}
	\call_|MMV| &\triangleq
	- \ln \prod_{g=1}^G p(\by^{(g)}|\bz,\btheta;\beta,\bgamma)
	p(\bz;\zeta) p(\btheta) + \const, \\
	&= \sum_{g=1}^G \left[
		\ln |\bC| + \left(\by^{(g)}\right)\h\bC\ii\by^{(g)} \right] \\
	& \hspace{10mm}
		- \sum_{k=1}^{K_|max|} \left( z_k\ln\zeta + (1-z_k)\ln(1-\zeta)\right)
		+ \const,
\end{align*}
where $p(\by^{(g)}|\bz,\btheta;\beta,\bgamma)=\CN(\by^{(g)}; \bm0,\bC)$
with \bC as in \eqref{py}. The posterior probabilities of the coefficient
vectors $\balpha^{(g)}$, $g=1,\ldots,G$, are given by \eqref{palpha} with $\by$ and $\balpha$
replaced by $\by^{(g)}$ and $\balpha^{(g)}$.

The procedure to estimate the variables $\btheta$, $\bz$, $\bgamma$, $\beta$
and $\zeta$ follows straightforwardly from the method used in the SMV case.
Here we provide a brief discussion of the derivation of the update equations;
refer to Sec. \ref{sec:alg} for details.

To estimate $\btheta$ and $\bgamma$ the first- and second-order derivatives of $\call_|MMV|$
are needed. Denote the derivative \eqref{derivative} with $\by$ replaced by
$\by^{(g)}$ as $\frac{\partial\call^{(g)}}{\partial [\btheta_{\hat\bz}]_k}$.
Then we have
\begin{align*}
	\frac{\partial\call_|MMV|}{\partial [\btheta_{\hat\bz}]_k} =
	\sum_{g=1}^G
	\frac{\partial\call^{(g)}}{\partial [\btheta_{\hat\bz}]_k}.
\end{align*}
A similar result follows for the second-order derivative and the derivatives
with respect to $\bgamma_{\hat\bz}$.

The estimate of $\zeta$ is unchanged from the SMV case \eqref{zetaUpdate}.

To estimate the noise variance $\beta$, we write an upper bound of the same
form as \eqref{bound_explicit} and find its minimizer to be
\begin{multline*}
	\hat\beta = \max\!\Big(\varepsilon_\beta,
		M\ii\tr\big(\hat\bSigma\bA\h(\hat\btheta_{\hat\bz})\bA(\hat\btheta_{\hat\bz})\big)
		\\
	+ (GM)\ii \sum_{g=1}^G ||\by^{(g)} - \bA(\hat\btheta_{\hat\bz}) \hat\bmu^{(g)}||^2
	\Big),
\end{multline*}
where $\hat\bmu^{(g)}$ is given by \eqref{bmu} with $\by$ replaced by
$\by^{(g)}$.

To write the activation and deactivation criteria for the MMV model we rewrite
the objective in terms of the parameters of a single component, analogously to
\eqref{singlecomp}:
\begin{multline*}
	\call_|MMV|(z_k, \theta_k, \gamma_k)
	= z_k \Bigg( G\ln\!\big( 1 + \gamma_k s_{\sim k}(\theta_k)\big)
	\\
	- \sum_{g=1}^G \frac{|q_{\sim k}^{(g)}(\theta_k)|^2}{\gamma_k\ii +
	s_{\sim k}(\theta_k)}
	+ \ln\!\left(\frac{1-\hat\zeta}{\hat\zeta}\right) \Bigg) + \const,
	\shownumber \label{singlecomp-mmv}
\end{multline*}
where $q_{\sim k}^{(g)}(\theta_k)$ is given by \eqref{qTilde} with $\by$
replaced by $\by^{(g)}$.
We omit the details of the activation and deactivation stages, as they follow
straightforwardly from \eqref{singlecomp-mmv} and the description in Secs.
\ref{sec:deactivate} and \ref{sec:activate}.

The insightful reader may have noticed that the calculations required for
MMV are very similar to those required for SVM. In particular, the
matrix $\hat\bC$ is unchanged and the methods for calculating
matrix-vector products involving $\hat\bC\ii$ presented in Secs.
\ref{sec:superfast} and \ref{sec:semifast} can be utilized. All expressions
involving $\by$ (i.e., $\bq$, $\br$, $\bu$, $\bq^\calg$ and $\by\hat\bC\ii\by$)
must be calculated for each observation vector $\by^{(g)}$. This means that in
the case of complete observations, the generalized Schur algorithm can be used
so that the MMV algorithm has per-iteration complexity $\calo(N\log^2N + GN\log
N)$. With incomplete observations the semifast method can be used with
per-iteration complexity $\calo(N\hat K^2 + GN\log N)$.

\section{Experiments}
\label{sec:experiments}
\pgfplotsset{custom_axis_style1/.style={
	title style={font=\footnotesize},
	legend columns=1,
	legend style={font=\scriptsize},
	legend style={inner xsep=2pt, inner ysep=1pt, nodes={inner sep=0.8pt}},
	legend style={/tikz/every even column/.append style={column sep=2pt}},
	label style={font=\footnotesize},
	xlabel shift=-3pt,
	ylabel shift=-3pt,
	xticklabel style={font=\footnotesize},
	yticklabel style={font=\footnotesize},
	every axis plot/.append style={line width=1pt}
}} 
\pgfplotsset{custom_axis_style2/.style={
	yticklabel style={rotate=90}
}}

\subsection{Setup, Algorithms \& Metrics}
\label{sec:numerical_setup}
In our experiments we use the signal model \eqref{model}. In the following the
wrap-around distance on $[0,1)$ is used for all differences of frequencies (see
Footnote \ref{distance}). Unless otherwise noted, the
true frequencies are drawn randomly, such that the minimum
separation between any two frequencies is $2/N$. Specifically,
the frequencies are generated sequentially for $k=1,\ldots,K$ with the $k$th
frequency, $\tilde\theta_k$, drawn from a uniform distribution on the set
$\{\theta \in [0,1) : d(\theta, \tilde\theta_l) > 2/N \textrm{ for all } l<k \}$.

The true coefficients in $\tilde\balpha$ are generated i.i.d. random, with each
entry drawn as follows. First a circularly-symmetric complex normal random
variable $a_k$ with standard deviation $0.8$ is drawn. The coefficient is then
found as $\tilde\alpha_k = a_k + 0.2\,e^{j\arg(a_k)}$. The resulting random
variable has the property $|\tilde\alpha_k|\ge0.2$, i.e., all components have
significant magnitude. We use this specification to ensure that all components
can be distinguished from noise. After generating the set of $K$
frequencies and coefficients, the noise variance $\beta$ is selected such that
the desired signal-to-noise ratio (SNR) is obtained, with $\mathrm{SNR} \triangleq
||\bPhi\bPsi(\tilde\btheta)\tilde\balpha||^2 / (M\beta)$.

We compare the superfast LSE algorithm%
\footnote{We have published our code at
github.com/thomaslundgaard/superfast-lse. It is based on our own implementation
of the generalized Schur algorithm and the NUFFT available at
cims.nyu.edu/cmcl/nufft/nufft.html.}
with the following reference algorithms: variational Bayesian line spectral
estimation (VALSE) \cite{badiu-valse};
atomic soft thresholding%
\footnote{The code is available online at github.com/badrinarayan/astlinespec.}
(AST) \cite{bhaskar-atomic};
gridless SPICE%
\footnote{The code has kindly been provided by the authors.}
(GLS) \cite{yang-gridless};
ESPRIT \cite{roy-esprit,kung-statespace};
and a gridded solution obtained with the least absolute shrinkage and selection
operator (LASSO) solved using SpaRSA%
\footnote{The code is available online at lx.it.pt/$\sim$mtf/SpaRSA.}
\cite{wright-sparsa}.

The solution to the primal problem of AST \cite{bhaskar-atomic} directly
provides an estimate of the signal vector $\bh=\bPsi(\tilde\btheta)\tilde\balpha$.
This solution is known to be biased towards the all-zero solution (as is also
the case with the classical LASSO solution). A so-called \textit{debiased}
solution can be obtained by recovering the frequencies from the AST dual and
estimating the coefficients $\tilde\balpha$ via least-squares. As in
\cite{bhaskar-atomic}, we report here the debiased solution.
If the frequencies are separated by at least $2/N$, the AST algorithm is known
to exactly recover the frequencies in the noise-free case
\cite{tang-offthegrid,bhaskar-atomic,candes-superresolution}. In the noisy case
no such recovery guarantee exists, but a bound on the estimation error of the
signal vector $\bh$ is known \cite{bhaskar-atomic,candes-superresolution}. Unfortunately
this error bound does not apply to the debiased solution we report herein.

We use the variant of GLS \cite{yang-gridless} which uses SORTE \cite{he-sorte}
for model order estimation and MUSIC \cite{schmidt-music} for frequency
estimation.

ESPRIT requires an estimate of the signal covariance matrix and of the
model order. The former is obtained as the averaged sample covariance matrix
computed from the signal vector split into $N/3$ signal vectors of length
$2N/N$ using forward-backward smoothing. The model order is estimated with SORTE
\cite{he-sorte}.

The LASSO solution is obtained using a grid of size $8N$. We have observed that no
improvement in performance is achieved with a finer grid. The regularization
parameter of LASSO is selected as proposed in \cite{bhaskar-atomic} with
knowledge of the true noise variance. We use the debiased solution
returned by the SpaRSA solver.

In the evaluation of the signal reconstruction we have also
included an oracle estimator (denoted Oracle) which obtains a least squares
solution for $\tilde\balpha$ with known $\tilde\btheta$.

Three performance metrics are used: normalized mean-squared
error (NMSE) of the reconstructed signal, block success rate (BSR) and
component success rate (CSR). The NMSE is defined as
\begin{align*}
	\textrm{NMSE} \triangleq \frac{||\bPsi(\hat\btau)\hat\balpha
		- \bPsi(\tilde\btau)\tilde\balpha ||^2
		}{||\bPsi(\tilde\btau)\tilde\balpha||^2}.
\end{align*}

The BSR is the proportion of Monte Carlo trials in which the frequency vector
$\tilde\btheta$ is successfully recovered. Successful recovery is understood as
correct estimation of the model order $K$ and that
$||d(\tilde\btheta,\hat\btheta)||_\infty < 0.5/N$. The association of the
entries in $\hat\btheta$ to those in $\tilde\btheta$ is obtained by using the
Hungarian method \cite{kuhn-hungarian} (also known as Munkres assigment
algorithm) minimizing $||d(\tilde\btheta,\hat\btheta)||_2^2$.

The BSR can be misleading, since a trial is considered to be unsuccessful if
just a single component is misestimated; for example if a component is
represented in the estimate by two components with very close frequencies. We
therefore introduce the CSR, defined as follows:
\begin{align*}
	\mathrm{CSR} \triangleq \frac{
		\sum_{k=1}^{\hat K} S(\hat\theta_k, \tilde\btheta) + 
		\sum_{k=1}^{K} S(\tilde\theta_k, \hat\btheta)
		}{\hat K + K}
\end{align*}
with the success function $S(x,\ba)\triangleq\mathbb{1}[\minn{k} d(x,a_k) <
0.5/N]$, where $\mathbb{1}[\cdot]$ denotes the indicator function.
The reported CSR is averaged over a number of Monte Carlo trials. The CSR takes
values in $[0,1]$. A CSR of $1$ is achieved if, and only if, all estimated
components are in the vicinity of one or more true components and all true
components are in the vicinity of one or more estimated components.

\setlength\figureheight{28mm}
\setlength\figurewidth{70mm}
\begin{figure}[t]
	\centering
%
\begin{tikzpicture}

\begin{axis}[%
width=0.951\figurewidth,
height=\figureheight,
at={(0\figurewidth,0\figureheight)},
scale only axis,
xmin=-8,
xmax=10,
xlabel style={font=\color{white!15!black}},
xlabel={$\delta$},
ymin=0,
ymax=600,
ylabel style={font=\color{white!15!black}},
ylabel={Number of occurences},
axis background/.style={fill=white},
xmajorgrids,
ymajorgrids,
legend style={legend cell align=left, align=left, draw=white!15!black},
custom_axis_style1, custom_axis_style2
]
\addplot [color=red, line width=1.0pt]
  table[row sep=crcr]{%
-4	1\\
-3.8	1\\
-3.8	0\\
-3.6	0\\
-3.6	2\\
-3.4	2\\
-3.4	3\\
-3.2	3\\
-3.2	1\\
-3	1\\
-3	2\\
-2.8	2\\
-2.8	1\\
-2.6	1\\
-2.6	1\\
-2.4	1\\
-2.4	0\\
-2.2	0\\
-2.2	1\\
-2	1\\
-2	2\\
-1.8	2\\
-1.8	4\\
-1.6	4\\
-1.6	3\\
-1.4	3\\
-1.4	12\\
-1.2	12\\
-1.2	20\\
-1	20\\
-1	59\\
-0.8	59\\
-0.8	102\\
-0.6	102\\
-0.6	195\\
-0.4	195\\
-0.4	287\\
-0.2	287\\
-0.2	298\\
0	298\\
0	389\\
0.2	389\\
0.2	410\\
0.4	410\\
0.4	405\\
0.600000000000001	405\\
0.600000000000001	409\\
0.800000000000001	409\\
0.800000000000001	359\\
1	359\\
1	324\\
1.2	324\\
1.2	310\\
1.4	310\\
1.4	226\\
1.6	226\\
1.6	214\\
1.8	214\\
1.8	179\\
2	179\\
2	143\\
2.2	143\\
2.2	125\\
2.4	125\\
2.4	112\\
2.6	112\\
2.6	96\\
2.8	96\\
2.8	62\\
3	62\\
3	44\\
3.2	44\\
3.2	34\\
3.4	34\\
3.4	29\\
3.6	29\\
3.6	27\\
3.8	27\\
3.8	28\\
4	28\\
4	14\\
4.2	14\\
4.2	11\\
4.4	11\\
4.4	8\\
4.6	8\\
4.6	9\\
4.8	9\\
4.8	9\\
5	9\\
5	5\\
5.2	5\\
5.2	4\\
5.4	4\\
5.4	2\\
5.6	2\\
5.6	8\\
5.8	8\\
5.8	1\\
6	1\\
6	5\\
6.2	5\\
6.2	1\\
6.4	1\\
6.4	1\\
6.6	1\\
6.6	0\\
6.8	0\\
6.8	0\\
7	0\\
7	0\\
7.2	0\\
7.2	1\\
7.4	1\\
7.4	1\\
7.6	1\\
7.6	0\\
7.8	0\\
7.8	0\\
8	0\\
8	0\\
8.2	0\\
8.2	0\\
8.4	0\\
8.4	1\\
8.6	1\\
};
\addlegendentry{SNR=12dB}

\addplot [color=blue, dashed, line width=1.0pt]
  table[row sep=crcr]{%
-6.9	2\\
-6.6	2\\
-6.6	0\\
-6.3	0\\
-6.3	1\\
-6	1\\
-6	1\\
-5.7	1\\
-5.7	0\\
-5.4	0\\
-5.4	1\\
-5.1	1\\
-5.1	5\\
-4.8	5\\
-4.8	14\\
-4.5	14\\
-4.5	15\\
-4.2	15\\
-4.2	20\\
-3.9	20\\
-3.9	39\\
-3.6	39\\
-3.6	70\\
-3.3	70\\
-3.3	103\\
-3	103\\
-3	101\\
-2.7	101\\
-2.7	87\\
-2.4	87\\
-2.4	46\\
-2.1	46\\
-2.1	76\\
-1.8	76\\
-1.8	102\\
-1.5	102\\
-1.5	185\\
-1.2	185\\
-1.2	387\\
-0.9	387\\
-0.9	534\\
-0.600000000000001	534\\
-0.600000000000001	583\\
-0.3	583\\
-0.3	498\\
0	498\\
0	427\\
0.3	427\\
0.3	375\\
0.6	375\\
0.6	302\\
0.899999999999999	302\\
0.899999999999999	232\\
1.2	232\\
1.2	187\\
1.5	187\\
1.5	152\\
1.8	152\\
1.8	105\\
2.1	105\\
2.1	84\\
2.4	84\\
2.4	62\\
2.7	62\\
2.7	44\\
3	44\\
3	35\\
3.3	35\\
3.3	36\\
3.6	36\\
3.6	17\\
3.9	17\\
3.9	17\\
4.2	17\\
4.2	18\\
4.5	18\\
4.5	14\\
4.8	14\\
4.8	5\\
5.1	5\\
5.1	2\\
5.4	2\\
5.4	5\\
5.7	5\\
5.7	3\\
6	3\\
6	3\\
6.3	3\\
6.3	0\\
6.6	0\\
6.6	2\\
6.9	2\\
6.9	2\\
7.2	2\\
7.2	0\\
7.5	0\\
7.5	0\\
7.8	0\\
7.8	0\\
8.1	0\\
8.1	0\\
8.4	0\\
8.4	1\\
8.7	1\\
};
\addlegendentry{SNR=20dB}

\addplot [color=black, dotted, line width=1.0pt]
  table[row sep=crcr]{%
-6.9	1\\
-6.6	1\\
-6.6	0\\
-6.3	0\\
-6.3	2\\
-6	2\\
-6	2\\
-5.7	2\\
-5.7	3\\
-5.4	3\\
-5.4	7\\
-5.1	7\\
-5.1	11\\
-4.8	11\\
-4.8	10\\
-4.5	10\\
-4.5	24\\
-4.2	24\\
-4.2	39\\
-3.9	39\\
-3.9	74\\
-3.6	74\\
-3.6	114\\
-3.3	114\\
-3.3	179\\
-3	179\\
-3	303\\
-2.7	303\\
-2.7	472\\
-2.4	472\\
-2.4	471\\
-2.1	471\\
-2.1	222\\
-1.8	222\\
-1.8	230\\
-1.5	230\\
-1.5	292\\
-1.2	292\\
-1.2	498\\
-0.9	498\\
-0.9	485\\
-0.600000000000001	485\\
-0.600000000000001	417\\
-0.3	417\\
-0.3	296\\
0	296\\
0	235\\
0.3	235\\
0.3	179\\
0.6	179\\
0.6	114\\
0.899999999999999	114\\
0.899999999999999	101\\
1.2	101\\
1.2	59\\
1.5	59\\
1.5	41\\
1.8	41\\
1.8	34\\
2.1	34\\
2.1	20\\
2.4	20\\
2.4	22\\
2.7	22\\
2.7	10\\
3	10\\
3	5\\
3.3	5\\
3.3	7\\
3.6	7\\
3.6	6\\
3.9	6\\
3.9	4\\
4.2	4\\
4.2	3\\
4.5	3\\
4.5	4\\
4.8	4\\
4.8	1\\
5.1	1\\
5.1	1\\
5.4	1\\
5.4	1\\
5.7	1\\
5.7	0\\
6	0\\
6	0\\
6.3	0\\
6.3	0\\
6.6	0\\
6.6	0\\
6.9	0\\
6.9	0\\
7.2	0\\
7.2	0\\
7.5	0\\
7.5	0\\
7.8	0\\
7.8	1\\
8.1	1\\
};
\addlegendentry{SNR=30dB}

\end{axis}
\end{tikzpicture}%
	\vspace{-3mm}
	\caption{Histograms of $\delta$ values for different signal-to-noise ratios.
	Cases where $\delta>0$ correspond to cases where an artefact component is
	activated if the criterion \eqref{activate_derived} is used.}
	\label{fig:criterion}
\end{figure}
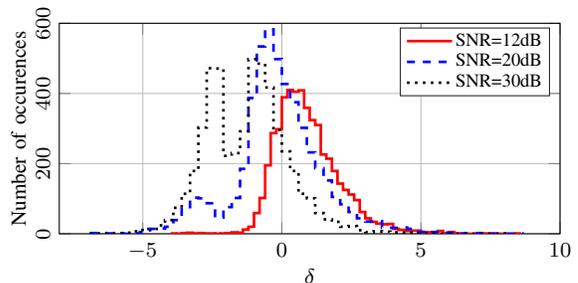

\subsection{Choosing the Activation Threshold}
\label{sec:threshold}
To determine a sensible value for the activation threshold $\tau$
in \eqref{activate_criterion}, the following experiment is conducted. We
consider the complete data case with $N=M=128$ and the number of components is
fixed at $K=35$, as there is a larger tendency to activate artefact components for
relatively large $K/N$. The algorithm is provided with the knowledge of $K_|max|=35$
and the activation probability is fixed at $\hat\zeta=35/128$. The algorithm is
run with the activation criterion \eqref{activate_derived}. In this way, the
algorithm in most cases successfully estimates the frequencies without any
artefacts.
After the algorithm has terminated, we test if $\tilde\btheta$ was successfully
recovered (as defined above). If so, $K_|max|$ is increased and the procedure
for activating a component in Sec. \ref{sec:activate} is run and the difference
between the left-hand and right-hand sides of \eqref{activate_derived} is saved.
We refer to this difference as $\delta$ and the criterion
\eqref{activate_derived} can be expressed as $\delta>0$. In Fig.
\ref{fig:criterion} we show histograms of the value $\delta$ obtained from
$5,000$ successful recoveries at three different SNR values At each SNR, the
experiment is repeated until the required number of successful recoveries are
obtained; trials without successful recovery are discarded. Cases where
$\delta>0$ thus correspond to cases where an artefact would be activated using
criterion \eqref{activate_derived}.

The heuristic criterion \eqref{activate_criterion} corresponds to $\delta>\tau$.
From Fig. \ref{fig:criterion} it is clearly seen that threshold $\tau=5$ is a
sensible value, which precludes almost all artefact components from being
activated. It is seen that this threshold works well for a large range
of SNR values. We have not investigated whether $\tau=5$ is
so large that desired components are precluded from activation.
The results in the following investigations are all obtained with $\tau=5$ and
the good performance of our algorithm across this wide selection of scenarios
indicates that the selected $\tau$ is not too large.

\setlength\figureheight{28mm}
\setlength\figurewidth{53mm}
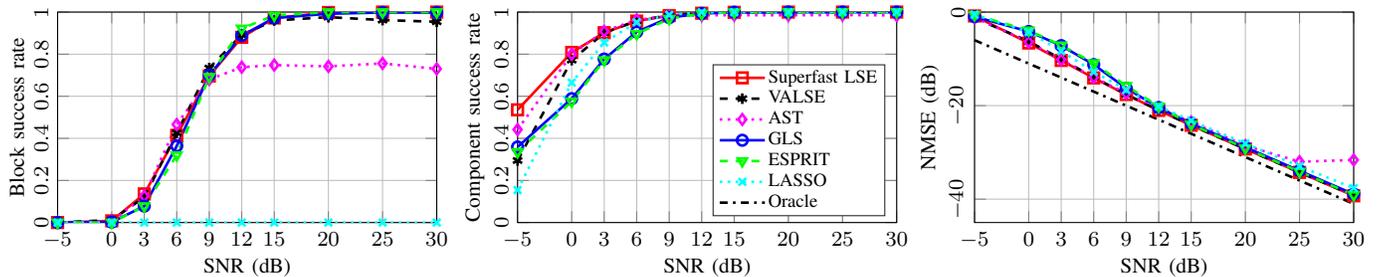
\begin{figure*}[t]
	\begin{minipage}[t]{0.334\linewidth}
		\centering
%
\definecolor{mycolor1}{rgb}{1.00000,0.00000,1.00000}%
\definecolor{mycolor2}{rgb}{0.00000,1.00000,1.00000}%
\begin{tikzpicture}

\begin{axis}[%
width=0.951\figurewidth,
height=\figureheight,
at={(0\figurewidth,0\figureheight)},
scale only axis,
xmin=-5,
xmax=30,
xlabel style={font=\color{white!15!black}},
xlabel={SNR (dB)},
ymin=0,
ymax=1,
ylabel style={font=\color{white!15!black}},
ylabel={Block success rate},
axis background/.style={fill=white},
xmajorgrids,
ymajorgrids,
custom_axis_style1, custom_axis_style2,
xtick = {-5.0000e+00,0.0000e+00,3.0000e+00,6.0000e+00,9.0000e+00,1.2000e+01,1.5000e+01,2.0000e+01,2.5000e+01,3.0000e+01}
]
\addplot [color=red, line width=1.0pt, mark=square, mark options={solid, red}, forget plot]
  table[row sep=crcr]{%
-5	0\\
0	0.008\\
3	0.136\\
6	0.414\\
9	0.696\\
12	0.88\\
15	0.972\\
20	0.998\\
25	1\\
30	1\\
};
\addplot [color=black, dashed, line width=1.0pt, mark=asterisk, mark options={solid, black}, forget plot]
  table[row sep=crcr]{%
-5	0\\
0	0.012\\
3	0.12\\
6	0.426\\
9	0.734\\
12	0.892\\
15	0.96\\
20	0.976\\
25	0.962\\
30	0.954\\
};
\addplot [color=mycolor1, dotted, line width=1.0pt, mark=diamond, mark options={solid, mycolor1}, forget plot]
  table[row sep=crcr]{%
-5	0\\
0	0.006\\
3	0.126\\
6	0.466\\
9	0.68\\
12	0.738\\
15	0.748\\
20	0.742\\
25	0.756\\
30	0.73\\
};
\addplot [color=blue, line width=1.0pt, mark=o, mark options={solid, blue}, forget plot]
  table[row sep=crcr]{%
-5	0\\
0	0.002\\
3	0.078\\
6	0.364\\
9	0.7\\
12	0.888\\
15	0.972\\
20	0.992\\
25	0.998\\
30	0.998\\
};
\addplot [color=green, dashed, line width=1.0pt, mark=triangle, mark options={solid, rotate=180, green}, forget plot]
  table[row sep=crcr]{%
-5	0\\
0	0\\
3	0.078\\
6	0.32\\
9	0.692\\
12	0.922\\
15	0.984\\
20	1\\
25	1\\
30	1\\
};
\addplot [color=mycolor2, dotted, line width=1.0pt, mark=x, mark options={solid, mycolor2}, forget plot]
  table[row sep=crcr]{%
-5	0\\
0	0\\
3	0\\
6	0\\
9	0\\
12	0\\
15	0\\
20	0\\
25	0\\
30	0\\
};
\addplot [color=black, dashdotted, line width=1.0pt, forget plot]
  table[row sep=crcr]{%
-1	0\\
};
\end{axis}
\end{tikzpicture}%
	\end{minipage}%
	\begin{minipage}[t]{0.334\linewidth}
		\centering
%
\definecolor{mycolor1}{rgb}{1.00000,0.00000,1.00000}%
\definecolor{mycolor2}{rgb}{0.00000,1.00000,1.00000}%
\begin{tikzpicture}

\begin{axis}[%
width=0.951\figurewidth,
height=\figureheight,
at={(0\figurewidth,0\figureheight)},
scale only axis,
xmin=-5,
xmax=30,
xlabel style={font=\color{white!15!black}},
xlabel={SNR (dB)},
ymin=0,
ymax=1,
ylabel style={font=\color{white!15!black}},
ylabel={Component success rate},
axis background/.style={fill=white},
xmajorgrids,
ymajorgrids,
legend style={at={(0.97,0.03)}, anchor=south east, legend cell align=left, align=left, draw=white!15!black},
custom_axis_style1, custom_axis_style2,
xtick = {-5.0000e+00,0.0000e+00,3.0000e+00,6.0000e+00,9.0000e+00,1.2000e+01,1.5000e+01,2.0000e+01,2.5000e+01,3.0000e+01}
]
\addplot [color=red, line width=1.0pt, mark=square, mark options={solid, red}]
  table[row sep=crcr]{%
-5	0.535605436405898\\
0	0.808367328911503\\
3	0.902011329190945\\
6	0.957167903425155\\
9	0.983369838297987\\
12	0.994570061517431\\
15	0.998761904761905\\
20	0.999904761904762\\
25	1\\
30	1\\
};
\addlegendentry{Superfast LSE}

\addplot [color=black, dashed, line width=1.0pt, mark=asterisk, mark options={solid, black}]
  table[row sep=crcr]{%
-5	0.294965211637349\\
0	0.771793569276777\\
3	0.900965586402653\\
6	0.959052090060077\\
9	0.984792855656931\\
12	0.994796992481204\\
15	0.998370927318296\\
20	0.999718614718615\\
25	0.999904761904762\\
30	1\\
};
\addlegendentry{VALSE}

\addplot [color=mycolor1, dotted, line width=1.0pt, mark=diamond, mark options={solid, mycolor1}]
  table[row sep=crcr]{%
-5	0.441895545527744\\
0	0.799208681648291\\
3	0.909062000560578\\
6	0.96276491519429\\
9	0.98044679815481\\
12	0.984736735118431\\
15	0.985472802559761\\
20	0.985005270092229\\
25	0.986083192170151\\
30	0.984164577451536\\
};
\addlegendentry{AST}

\addplot [color=blue, line width=1.0pt, mark=o, mark options={solid, blue}]
  table[row sep=crcr]{%
-5	0.359094307301157\\
0	0.588843406316691\\
3	0.776754944651528\\
6	0.905564398511111\\
9	0.973544700043531\\
12	0.993644013107636\\
15	0.999275689223058\\
20	1\\
25	1\\
30	1\\
};
\addlegendentry{GLS}

\addplot [color=green, dashed, line width=1.0pt, mark=triangle, mark options={solid, rotate=180, green}]
  table[row sep=crcr]{%
-5	0.335439349572901\\
0	0.575190800656884\\
3	0.771485601260588\\
6	0.896747137972933\\
9	0.968111576255219\\
12	0.992891077962286\\
15	0.998948279790385\\
20	1\\
25	1\\
30	1\\
};
\addlegendentry{ESPRIT}

\addplot [color=mycolor2, dotted, line width=1.0pt, mark=x, mark options={solid, mycolor2}]
  table[row sep=crcr]{%
-5	0.154290226440227\\
0	0.664825409384069\\
3	0.855393417248461\\
6	0.951635493153343\\
9	0.987020252452861\\
12	0.998662439637657\\
15	0.999928571428571\\
20	0.999935483870968\\
25	0.999935483870968\\
30	0.999935483870968\\
};
\addlegendentry{LASSO}

\addplot [color=black, dashdotted, line width=1.0pt]
  table[row sep=crcr]{%
-1	0\\
};
\addlegendentry{Oracle}

\end{axis}
\end{tikzpicture}%
	\end{minipage}%
	\begin{minipage}[t]{0.334\linewidth}
		\centering
%
\definecolor{mycolor1}{rgb}{1.00000,0.00000,1.00000}%
\definecolor{mycolor2}{rgb}{0.00000,1.00000,1.00000}%
\begin{tikzpicture}

\begin{axis}[%
width=0.951\figurewidth,
height=\figureheight,
at={(0\figurewidth,0\figureheight)},
scale only axis,
xmin=-5,
xmax=30,
xlabel style={font=\color{white!15!black}},
xlabel={SNR (dB)},
ymin=-45,
ymax=0,
ylabel style={font=\color{white!15!black}},
ylabel={NMSE (dB)},
axis background/.style={fill=white},
xmajorgrids,
ymajorgrids,
custom_axis_style1, custom_axis_style2,
xtick = {-5.0000e+00,0.0000e+00,3.0000e+00,6.0000e+00,9.0000e+00,1.2000e+01,1.5000e+01,2.0000e+01,2.5000e+01,3.0000e+01}
]
\addplot [color=red, line width=1.0pt, mark=square, mark options={solid, red}, forget plot]
  table[row sep=crcr]{%
-5	-0.825152569174306\\
0	-6.61246401679639\\
3	-10.3364668047529\\
6	-14.0789070633925\\
9	-17.6245969316508\\
12	-21.0176865807834\\
15	-24.201812038367\\
20	-29.2700982564923\\
25	-34.2795999141231\\
30	-39.2813114209182\\
};
\addplot [color=black, dashed, line width=1.0pt, mark=asterisk, mark options={solid, black}, forget plot]
  table[row sep=crcr]{%
-5	-1.27781962563817\\
0	-6.23672943636677\\
3	-10.0588629153294\\
6	-13.9429330475816\\
9	-17.5949552770815\\
12	-21.0033829970373\\
15	-24.1766078731784\\
20	-29.2556330383114\\
25	-34.2565140729674\\
30	-39.2044417585279\\
};
\addplot [color=mycolor1, dotted, line width=1.0pt, mark=diamond, mark options={solid, mycolor1}, forget plot]
  table[row sep=crcr]{%
-5	-1.18127690767484\\
0	-6.40203461017441\\
3	-10.1746300846156\\
6	-13.9574923670184\\
9	-17.3691166864654\\
12	-20.5042827412881\\
15	-23.5040812309196\\
20	-28.5488904314176\\
25	-31.9981907521871\\
30	-31.6225574682831\\
};
\addplot [color=blue, line width=1.0pt, mark=o, mark options={solid, blue}, forget plot]
  table[row sep=crcr]{%
-5	-0.974931390517965\\
0	-4.15120572407836\\
3	-7.17335503805815\\
6	-11.2357803842631\\
9	-16.2752053955644\\
12	-20.4211304163949\\
15	-23.8447260530212\\
20	-28.9053900205189\\
25	-33.9227156307388\\
30	-38.9291699451771\\
};
\addplot [color=green, dashed, line width=1.0pt, mark=triangle, mark options={solid, rotate=180, green}, forget plot]
  table[row sep=crcr]{%
-5	-0.603753641690292\\
0	-3.92045544891863\\
3	-7.11375996650937\\
6	-10.9098200718216\\
9	-15.7714765230702\\
12	-20.1663800267733\\
15	-24.019558696187\\
20	-29.109376412624\\
25	-34.1166587563532\\
30	-39.1183194926366\\
};
\addplot [color=mycolor2, dotted, line width=1.0pt, mark=x, mark options={solid, mycolor2}, forget plot]
  table[row sep=crcr]{%
-5	-0.534546006954481\\
0	-4.26491100582623\\
3	-8.3450084337756\\
6	-12.732897024072\\
9	-16.6612341563831\\
12	-20.1604443273287\\
15	-23.2115337423787\\
20	-28.0922847544193\\
25	-32.9697836116137\\
30	-37.6401331900758\\
};
\addplot [color=black, dashdotted, line width=1.0pt, forget plot]
  table[row sep=crcr]{%
-5	-6.00448306762345\\
0	-11.0044830676235\\
3	-14.0044830676235\\
6	-17.0044830676235\\
9	-20.0044830676235\\
12	-23.0044830676235\\
15	-26.0044830676235\\
20	-31.0044830676235\\
25	-36.0044830676235\\
30	-41.0044830676235\\
};
\end{axis}
\end{tikzpicture}%
	\end{minipage}%
	\vspace{-3mm}
	\caption{Simulation results for varying SNR with complete data
	($\bPhi=\bI$). The signal length is $N=M=128$ and the number of components
	is $K=10$.  Results are averaged over $500$ Monte Carlo trials. The legend
	applies to all plots. Only the NMSE of Oracle is shown.}
	\label{fig:snr}
\end{figure*}

\pgfplotsset{custom_axis_style2/.append style={
	xticklabel style={
		rotate=45,
		/pgf/number format/.cd,
		fixed,
		fixed zerofill,
		precision=2,
		/tikz/.cd,
		yshift=3pt
	}
}} 
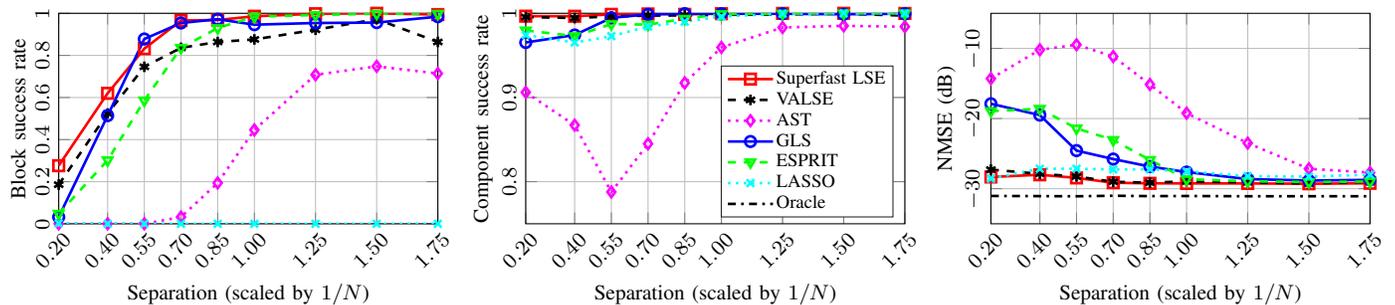
\begin{figure*}[t]
	\begin{minipage}[t]{0.34\linewidth}
		\centering
%
\definecolor{mycolor1}{rgb}{1.00000,0.00000,1.00000}%
\definecolor{mycolor2}{rgb}{0.00000,1.00000,1.00000}%
\begin{tikzpicture}

\begin{axis}[%
width=0.951\figurewidth,
height=\figureheight,
at={(0\figurewidth,0\figureheight)},
scale only axis,
xmin=0.2,
xmax=1.75,
xlabel style={font=\color{white!15!black}},
xlabel={Separation (scaled by $1/N$)},
ymin=0,
ymax=1,
ylabel style={font=\color{white!15!black}},
ylabel={Block success rate},
axis background/.style={fill=white},
xmajorgrids,
ymajorgrids,
custom_axis_style1, custom_axis_style2,
xtick = {2.0000e-01,4.0000e-01,5.5000e-01,7.0000e-01,8.5000e-01,1.0000e+00,1.2500e+00,1.5000e+00,1.7500e+00}
]
\addplot [color=red, line width=1.0pt, mark=square, mark options={solid, red}, forget plot]
  table[row sep=crcr]{%
0.2	0.276\\
0.4	0.62\\
0.55	0.832\\
0.7	0.966\\
0.85	0.968\\
1	0.986\\
1.25	0.998\\
1.5	1\\
1.75	0.994\\
};
\addplot [color=black, dashed, line width=1.0pt, mark=asterisk, mark options={solid, black}, forget plot]
  table[row sep=crcr]{%
0.2	0.186\\
0.4	0.524\\
0.55	0.746\\
0.7	0.836\\
0.85	0.864\\
1	0.876\\
1.25	0.922\\
1.5	0.972\\
1.75	0.864\\
};
\addplot [color=mycolor1, dotted, line width=1.0pt, mark=diamond, mark options={solid, mycolor1}, forget plot]
  table[row sep=crcr]{%
0.2	0\\
0.4	0\\
0.55	0\\
0.7	0.032\\
0.85	0.194\\
1	0.446\\
1.25	0.708\\
1.5	0.748\\
1.75	0.714\\
};
\addplot [color=blue, line width=1.0pt, mark=o, mark options={solid, blue}, forget plot]
  table[row sep=crcr]{%
0.2	0.032\\
0.4	0.514\\
0.55	0.878\\
0.7	0.954\\
0.85	0.972\\
1	0.946\\
1.25	0.954\\
1.5	0.956\\
1.75	0.984\\
};
\addplot [color=green, dashed, line width=1.0pt, mark=triangle, mark options={solid, rotate=180, green}, forget plot]
  table[row sep=crcr]{%
0.2	0.048\\
0.4	0.302\\
0.55	0.586\\
0.7	0.836\\
0.85	0.93\\
1	0.984\\
1.25	0.992\\
1.5	0.998\\
1.75	0.998\\
};
\addplot [color=mycolor2, dotted, line width=1.0pt, mark=x, mark options={solid, mycolor2}, forget plot]
  table[row sep=crcr]{%
0.2	0.002\\
0.4	0\\
0.55	0\\
0.7	0\\
0.85	0\\
1	0\\
1.25	0\\
1.5	0\\
1.75	0\\
};
\addplot [color=black, dashdotted, line width=1.0pt, forget plot]
  table[row sep=crcr]{%
-1	0\\
};
\end{axis}
\end{tikzpicture}%
	\end{minipage}%
	\begin{minipage}[t]{0.34\linewidth}
		\centering
%
\definecolor{mycolor1}{rgb}{1.00000,0.00000,1.00000}%
\definecolor{mycolor2}{rgb}{0.00000,1.00000,1.00000}%
\begin{tikzpicture}

\begin{axis}[%
width=0.951\figurewidth,
height=\figureheight,
at={(0\figurewidth,0\figureheight)},
scale only axis,
xmin=0.2,
xmax=1.75,
xlabel style={font=\color{white!15!black}},
xlabel={Separation (scaled by $1/N$)},
ymin=0.75,
ymax=1,
ylabel style={font=\color{white!15!black}},
ylabel={Component success rate},
axis background/.style={fill=white},
xmajorgrids,
ymajorgrids,
legend style={at={(0.97,0.03)}, anchor=south east, legend cell align=left, align=left, draw=white!15!black},
custom_axis_style1, custom_axis_style2,
xtick = {2.0000e-01,4.0000e-01,5.5000e-01,7.0000e-01,8.5000e-01,1.0000e+00,1.2500e+00,1.5000e+00,1.7500e+00}
]
\addplot [color=red, line width=1.0pt, mark=square, mark options={solid, red}]
  table[row sep=crcr]{%
0.2	0.996596201287533\\
0.4	0.996512945241893\\
0.55	0.998722305764411\\
0.7	0.999218045112782\\
0.85	0.998761904761905\\
1	0.99952380952381\\
1.25	0.999904761904762\\
1.5	1\\
1.75	0.999809523809524\\
};
\addlegendentry{Superfast LSE}

\addplot [color=black, dashed, line width=1.0pt, mark=asterisk, mark options={solid, black}]
  table[row sep=crcr]{%
0.2	0.995507488864765\\
0.4	0.994421301422587\\
0.55	0.996302984734564\\
0.7	0.997244520391889\\
0.85	0.997471485036703\\
1	0.998151667046404\\
1.25	0.998669628616997\\
1.5	0.999528138528139\\
1.75	0.997235361130098\\
};
\addlegendentry{VALSE}

\addplot [color=mycolor1, dotted, line width=1.0pt, mark=diamond, mark options={solid, mycolor1}]
  table[row sep=crcr]{%
0.2	0.906122182840267\\
0.4	0.86712234956336\\
0.55	0.787773591712667\\
0.7	0.84500196585148\\
0.85	0.917324010540348\\
1	0.959500378022693\\
1.25	0.983171582117351\\
1.5	0.985109944822532\\
1.75	0.984090315481622\\
};
\addlegendentry{AST}

\addplot [color=blue, line width=1.0pt, mark=o, mark options={solid, blue}]
  table[row sep=crcr]{%
0.2	0.965336849382127\\
0.4	0.974220589919943\\
0.55	0.994602230956179\\
0.7	0.998724462671831\\
0.85	0.999609022556391\\
1	0.999513784461153\\
1.25	0.999437229437229\\
1.5	0.998851446798815\\
1.75	0.999513784461153\\
};
\addlegendentry{GLS}

\addplot [color=green, dashed, line width=1.0pt, mark=triangle, mark options={solid, rotate=180, green}]
  table[row sep=crcr]{%
0.2	0.979653126500262\\
0.4	0.972925173519292\\
0.55	0.987184212631195\\
0.7	0.986577594799795\\
0.85	0.993519912148396\\
1	0.999071010860485\\
1.25	0.999619047619048\\
1.5	0.999904761904762\\
1.75	0.999904761904762\\
};
\addlegendentry{ESPRIT}

\addplot [color=mycolor2, dotted, line width=1.0pt, mark=x, mark options={solid, mycolor2}]
  table[row sep=crcr]{%
0.2	0.974209365312609\\
0.4	0.965356847419959\\
0.55	0.972815472961064\\
0.7	0.984306512670922\\
0.85	0.989966730034424\\
1	0.996137033138117\\
1.25	0.999543099776567\\
1.5	1\\
1.75	0.999864367816092\\
};
\addlegendentry{LASSO}

\addplot [color=black, dashdotted, line width=1.0pt]
  table[row sep=crcr]{%
-1	0\\
};
\addlegendentry{Oracle}

\end{axis}
\end{tikzpicture}%
	\end{minipage}%
	\begin{minipage}[t]{0.34\linewidth}
		\centering
%
\definecolor{mycolor1}{rgb}{1.00000,0.00000,1.00000}%
\definecolor{mycolor2}{rgb}{0.00000,1.00000,1.00000}%
\begin{tikzpicture}

\begin{axis}[%
width=0.951\figurewidth,
height=\figureheight,
at={(0\figurewidth,0\figureheight)},
scale only axis,
xmin=0.2,
xmax=1.75,
xlabel style={font=\color{white!15!black}},
xlabel={Separation (scaled by $1/N$)},
ymin=-35,
ymax=-5,
ylabel style={font=\color{white!15!black}},
ylabel={NMSE (dB)},
axis background/.style={fill=white},
xmajorgrids,
ymajorgrids,
custom_axis_style1, custom_axis_style2,
xtick = {2.0000e-01,4.0000e-01,5.5000e-01,7.0000e-01,8.5000e-01,1.0000e+00,1.2500e+00,1.5000e+00,1.7500e+00}
]
\addplot [color=red, line width=1.0pt, mark=square, mark options={solid, red}, forget plot]
  table[row sep=crcr]{%
0.2	-28.3313123707022\\
0.4	-28.0130584321394\\
0.55	-28.4630075961934\\
0.7	-29.152268553035\\
0.85	-29.2330706694602\\
1	-29.2475702948394\\
1.25	-29.2503016515383\\
1.5	-29.2947969461149\\
1.75	-29.2282079369465\\
};
\addplot [color=black, dashed, line width=1.0pt, mark=asterisk, mark options={solid, black}, forget plot]
  table[row sep=crcr]{%
0.2	-27.2810577548178\\
0.4	-27.8806785447787\\
0.55	-28.2601071506842\\
0.7	-29.0107432437488\\
0.85	-29.1433962011915\\
1	-28.952143409647\\
1.25	-29.0960278656082\\
1.5	-29.2581154631426\\
1.75	-29.0173316237517\\
};
\addplot [color=mycolor1, dotted, line width=1.0pt, mark=diamond, mark options={solid, mycolor1}, forget plot]
  table[row sep=crcr]{%
0.2	-14.309229995246\\
0.4	-10.2024757545123\\
0.55	-9.47993259521013\\
0.7	-11.1553746893641\\
0.85	-15.1420192841986\\
1	-19.2330123422758\\
1.25	-23.4973164189905\\
1.5	-27.1639350973654\\
1.75	-27.6966703071522\\
};
\addplot [color=blue, line width=1.0pt, mark=o, mark options={solid, blue}, forget plot]
  table[row sep=crcr]{%
0.2	-17.8903300412165\\
0.4	-19.5077557225192\\
0.55	-24.5688276280666\\
0.7	-25.775555862465\\
0.85	-26.8422925641442\\
1	-27.6316060574662\\
1.25	-28.6284361175796\\
1.5	-28.8383595275377\\
1.75	-28.7433861330341\\
};
\addplot [color=green, dashed, line width=1.0pt, mark=triangle, mark options={solid, rotate=180, green}, forget plot]
  table[row sep=crcr]{%
0.2	-18.9378306968805\\
0.4	-18.6136583315928\\
0.55	-21.4033655241915\\
0.7	-23.0544871691449\\
0.85	-25.9008507752883\\
1	-28.6512492469105\\
1.25	-28.9326553254776\\
1.5	-29.0509510046933\\
1.75	-28.9714577226897\\
};
\addplot [color=mycolor2, dotted, line width=1.0pt, mark=x, mark options={solid, mycolor2}, forget plot]
  table[row sep=crcr]{%
0.2	-28.5627173670501\\
0.4	-27.153546342416\\
0.55	-27.1524405267531\\
0.7	-27.2537491858861\\
0.85	-27.2311096084533\\
1	-27.545520196932\\
1.25	-28.2023461818086\\
1.5	-28.2670891615651\\
1.75	-27.9975783588323\\
};
\addplot [color=black, dashdotted, line width=1.0pt, forget plot]
  table[row sep=crcr]{%
0.2	-31.0332288249624\\
0.4	-31.0580333981046\\
0.55	-31.0815398223317\\
0.7	-31.003930512676\\
0.85	-31.0572176663486\\
1	-31.0429659315615\\
1.25	-31.0770494245551\\
1.5	-31.073403351987\\
1.75	-31.0761293657706\\
};
\end{axis}
\end{tikzpicture}%
	\end{minipage}%
	\vspace{-3mm}
	\caption{Simulation results for closely23 located components with complete
	data ($\bPhi=\bI$). The frequencies are generated as 5 pairs (i.e. $K=10$)
	such that each pair has varying (deterministic) intra-pair separation,
	while the location of the pairs are generated randomly with non-paired
	frequencies at least $2/N$ apart (i.e., the location of the pairs are
	generated using a procedure similar to the one described in Sec.
	\ref{sec:numerical_setup}). The signal length is $N=M=128$ and the SNR is
	$20\db$. Results are averaged over $500$ Monte Carlo trials. The legend
	applies to all plots. Only the NMSE of Oracle is shown.}
	\label{fig:pairsep}
\end{figure*}

\subsection{Estimation with Complete Data}
In \mbox{Fig.\@ \ref{fig:snr}} we show performance results versus SNR. We first
notice that Superfast LSE is on par with or outperforms all other algorithms in
the three metrics shown here for all SNR values. In the low SNR region no
algorithm can reliably recover the correct model order and the frequencies. In
the plots of the CSR and NMSE, we see that Superfast LSE, VALSE and AST
generally achieve the best approximation of the true frequencies. There is a
small performance gap in terms of NMSE between Oracle and all other evaluated
algorithms due to the uncertainty in frequency estimation (Oracle knows the
true frequencies).

ESPRIT and GLS are observed to have the weakest performance at low SNR,
especially in terms of CSR and NMSE. Both algorithms use SORTE to estimate the
model order from the eigenvalues of the signal covariance. At low SNR it is
hard to distinguish the signal eigenvalues from the noise eigenvalues, leading
to the observed deterioration in performance.

At medium to high SNR, BSR of AST is about $0.75$.
The algorithm tends to slightly overestimate the model order (not shown here).
We hypothesise that such systematic overestimation of the model order can be
avoided by adjusting the regularization parameter used in AST. Doing so would,
however, mean that AST would perform worse in other scenarios. This is exactly
the weakness of methods involving regularization parameters.

Finally note that LASSO is never able to successfully estimate the model order,
due to the use of a grid. In particular each true frequency component is estimated by a
few non-zero entries at neighbouring gridpoints. It is visible in the CSR that
LASSO indeed estimates frequencies which lie in the vicinity of the
true frequencies. In some applications, e.g. channel estimation in wireless
communications, it is the reconstructed signal and not
the frequencies themself which are of interest. In this case LASSO may be
preferable because of its simplicity. Due to the grid approximation, LASSO
performs a little worse than the best gridless algorithms in terms of
NMSE.


\pgfplotsset{custom_axis_style2/.style={
	yticklabel style={rotate=90}
}}
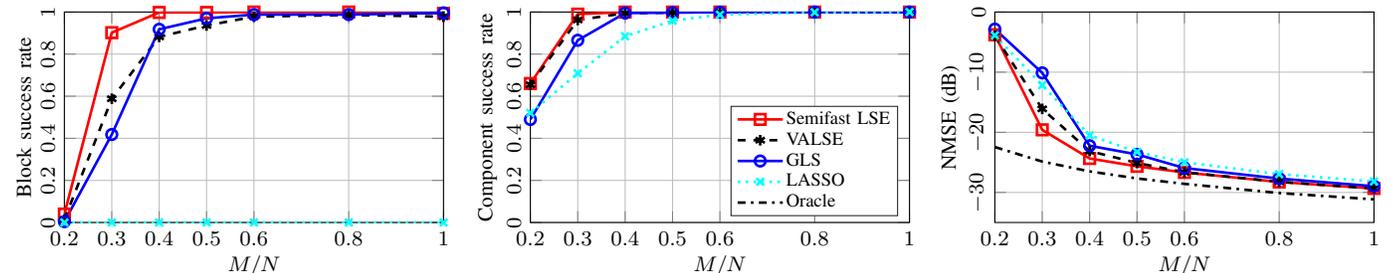
\begin{figure*}[t]
	\begin{minipage}[t]{0.34\linewidth}
		\centering
%
\definecolor{mycolor1}{rgb}{0.00000,1.00000,1.00000}%
\begin{tikzpicture}

\begin{axis}[%
width=0.951\figurewidth,
height=\figureheight,
at={(0\figurewidth,0\figureheight)},
scale only axis,
xmin=0.2,
xmax=1,
xlabel style={font=\color{white!15!black}},
xlabel={$M/N$},
ymin=0,
ymax=1,
ylabel style={font=\color{white!15!black}},
ylabel={Block success rate},
axis background/.style={fill=white},
xmajorgrids,
ymajorgrids,
custom_axis_style1, custom_axis_style2,
xtick = {2.0000e-01,3.0000e-01,4.0000e-01,5.0000e-01,6.0000e-01,8.0000e-01,1.0000e+00}
]
\addplot [color=red, line width=1.0pt, mark=square, mark options={solid, red}, forget plot]
  table[row sep=crcr]{%
0.2	0.04\\
0.3	0.902\\
0.4	0.998\\
0.5	0.998\\
0.6	1\\
0.8	1\\
1	0.994\\
};
\addplot [color=black, dashed, line width=1.0pt, mark=asterisk, mark options={solid, black}, forget plot]
  table[row sep=crcr]{%
0.2	0.014\\
0.3	0.588\\
0.4	0.882\\
0.5	0.936\\
0.6	0.978\\
0.8	0.986\\
1	0.978\\
};
\addplot [color=blue, line width=1.0pt, mark=o, mark options={solid, blue}, forget plot]
  table[row sep=crcr]{%
0.2	0.004\\
0.3	0.418\\
0.4	0.918\\
0.5	0.97\\
0.6	0.988\\
0.8	0.988\\
1	0.996\\
};
\addplot [color=mycolor1, dotted, line width=1.0pt, mark=x, mark options={solid, mycolor1}, forget plot]
  table[row sep=crcr]{%
0.2	0\\
0.3	0\\
0.4	0\\
0.5	0\\
0.6	0\\
0.8	0\\
1	0\\
};
\addplot [color=black, dashdotted, line width=1.0pt, forget plot]
  table[row sep=crcr]{%
-1	0\\
};
\end{axis}
\end{tikzpicture}%
	\end{minipage}%
	\begin{minipage}[t]{0.34\linewidth}
		\centering
%
\definecolor{mycolor1}{rgb}{0.00000,1.00000,1.00000}%
\begin{tikzpicture}

\begin{axis}[%
width=0.951\figurewidth,
height=\figureheight,
at={(0\figurewidth,0\figureheight)},
scale only axis,
xmin=0.2,
xmax=1,
xlabel style={font=\color{white!15!black}},
xlabel={$M/N$},
ymin=0,
ymax=1,
ylabel style={font=\color{white!15!black}},
ylabel={Component success rate},
axis background/.style={fill=white},
xmajorgrids,
ymajorgrids,
legend style={at={(0.97,0.03)}, anchor=south east, legend cell align=left, align=left, draw=white!15!black},
custom_axis_style1, custom_axis_style2,
xtick = {2.0000e-01,3.0000e-01,4.0000e-01,5.0000e-01,6.0000e-01,8.0000e-01,1.0000e+00}
]
\addplot [color=red, line width=1.0pt, mark=square, mark options={solid, red}]
  table[row sep=crcr]{%
0.2	0.660337112773558\\
0.3	0.989997260307633\\
0.4	0.999894736842105\\
0.5	0.999904761904762\\
0.6	1\\
0.8	1\\
1	0.999809523809524\\
};
\addlegendentry{Semifast LSE}

\addplot [color=black, dashed, line width=1.0pt, mark=asterisk, mark options={solid, black}]
  table[row sep=crcr]{%
0.2	0.654582869246337\\
0.3	0.962370450636399\\
0.4	0.99436243058627\\
0.5	0.997450292397661\\
0.6	0.999894736842105\\
0.8	0.999894736842105\\
1	0.999904761904762\\
};
\addlegendentry{VALSE}

\addplot [color=blue, line width=1.0pt, mark=o, mark options={solid, blue}]
  table[row sep=crcr]{%
0.2	0.489458479305801\\
0.3	0.86578681287703\\
0.4	0.99511403508772\\
0.5	0.99710883268778\\
0.6	0.99949373433584\\
0.8	0.999799498746867\\
1	0.999904761904762\\
};
\addlegendentry{GLS}

\addplot [color=mycolor1, dotted, line width=1.0pt, mark=x, mark options={solid, mycolor1}]
  table[row sep=crcr]{%
0.2	0.520670547949162\\
0.3	0.708584551737307\\
0.4	0.884607937477732\\
0.5	0.958856597195834\\
0.6	0.987259373001961\\
0.8	0.998676140155728\\
1	1\\
};
\addlegendentry{LASSO}

\addplot [color=black, dashdotted, line width=1.0pt]
  table[row sep=crcr]{%
-1	0\\
};
\addlegendentry{Oracle}

\end{axis}
\end{tikzpicture}%
	\end{minipage}%
	\begin{minipage}[t]{0.34\linewidth}
		\centering
%
\definecolor{mycolor1}{rgb}{0.00000,1.00000,1.00000}%
\begin{tikzpicture}

\begin{axis}[%
width=0.951\figurewidth,
height=\figureheight,
at={(0\figurewidth,0\figureheight)},
scale only axis,
xmin=0.2,
xmax=1,
xlabel style={font=\color{white!15!black}},
xlabel={$M/N$},
ymin=-35,
ymax=0,
ylabel style={font=\color{white!15!black}},
ylabel={NMSE (dB)},
axis background/.style={fill=white},
xmajorgrids,
ymajorgrids,
custom_axis_style1, custom_axis_style2,
xtick = {2.0000e-01,3.0000e-01,4.0000e-01,5.0000e-01,6.0000e-01,8.0000e-01,1.0000e+00}
]
\addplot [color=red, line width=1.0pt, mark=square, mark options={solid, red}, forget plot]
  table[row sep=crcr]{%
0.2	-3.77495945820166\\
0.3	-19.5909222416382\\
0.4	-24.3550223896901\\
0.5	-25.6585256942978\\
0.6	-26.6766816604897\\
0.8	-28.2601486925761\\
1	-29.3428154515451\\
};
\addplot [color=black, dashed, line width=1.0pt, mark=asterisk, mark options={solid, black}, forget plot]
  table[row sep=crcr]{%
0.2	-3.9609669813372\\
0.3	-16.1042490585478\\
0.4	-23.1184560357098\\
0.5	-25.1019937127937\\
0.6	-26.6437808784917\\
0.8	-28.2262592183048\\
1	-29.346074216869\\
};
\addplot [color=blue, line width=1.0pt, mark=o, mark options={solid, blue}, forget plot]
  table[row sep=crcr]{%
0.2	-2.8383112758973\\
0.3	-10.1141182916835\\
0.4	-22.2371566990053\\
0.5	-23.679420867998\\
0.6	-25.9462488366537\\
0.8	-27.6985259748107\\
1	-28.9666356706336\\
};
\addplot [color=mycolor1, dotted, line width=1.0pt, mark=x, mark options={solid, mycolor1}, forget plot]
  table[row sep=crcr]{%
0.2	-3.88228914589613\\
0.3	-12.1632727784128\\
0.4	-20.4994173413459\\
0.5	-23.2573406502259\\
0.6	-24.9919582468351\\
0.8	-26.9576855071278\\
1	-28.1744143226591\\
};
\addplot [color=black, dashdotted, line width=1.0pt, forget plot]
  table[row sep=crcr]{%
0.2	-22.450477541817\\
0.3	-24.8668138915482\\
0.4	-26.5003918411833\\
0.5	-27.6743045960145\\
0.6	-28.5737975753956\\
0.8	-30.1028844948305\\
1	-31.14784775835\\
};
\end{axis}
\end{tikzpicture}%
	\end{minipage}%
	\vspace{-3mm}
	\caption{Simulation results with incomplete data, i.e., $\bPhi$ contains $M$
	rows of $\bI$ selected at random.
	The signal length is $N=128$, the SNR is $20\db$ and
	the number of components is $K=10$.
	Results are averaged over $500$ Monte Carlo trials. The legend applies to
	all plots. Only the NMSE of Oracle is shown.}
	\label{fig:MtoN}
\end{figure*}

\subsection{Super Resolution}
The ability to separate components beyond the Rayleigh limit of $1/N$ is known
as super resolution. The results in Fig. \ref{fig:pairsep} illustrate the super
resolution ability of the algorithms. In this experiment we generate 5 pairs of
frequencies with varying distance between the two frequencies in each pair. The
pairs are well separated (at least $2/N$ separation between frequencies which
are not in the same pair).

The NMSE performance of Superfast LSE, VALSE and LASSO is only slightly worse
at low separation when compared to the performance at large separation. It is
evident that the model order and the frequencies cannot be recovered in every
case (BSR below 1) when the separation is less than $1/N$. Since the CSR is
close to 1 and the NMSE is close to that of Oracle, these three algorithms
handle closely located components well, in the sense that a good approximation
of the frequencies is obtained.

AST, GLS and ESPRIT give a CSR below 1 and a rather large NMSE when the
separation is small. This is despite the fact that GLS and ESPRIT do not show a
significantly worse BSR compared to Superfast LSE. We have observed that this
is because these algorithms significantly underestimate the model order in some
cases, resulting in a large contribution to NMSE.

ESPRIT shows worse super resolution ability than Superfast LSE, VALSE and GLS
(lower BSR for separation below $0.7/N$). This is because a covariance matrix
of size $2N/3$ is formed, thus reducing the effective signal length.

\setlength\figureheight{39mm}
\setlength\figurewidth{62mm}
\newcommand{\titletext}{bla}
\newcommand{\algname}{bla}
\pgfplotsset{custom_axis_style2/.style={
	xticklabel style = {rotate=90},
	yticklabel style={/pgf/number format/fixed}
}} 
\begin{figure*}[t]
	\captionsetup[sub]{margin={-10mm,0mm}}
	\begin{minipage}[t]{0.32\linewidth}
		\centering
		\renewcommand{\titletext}{Superfast LSE}
		\renewcommand{\algname}{SuperfastLSE}
		\pgfplotsset{colorbar=false}
		\subcaption{Superfast LSE}
		\begin{tikzpicture}
	\begin{axis}[
		width=0.951\figurewidth,
		height=\figureheight,
		at={(0\figurewidth,0\figureheight)},
		xlabel=Separation (scaled by $1/N$),
		ylabel=$K/N$,
		colormap={}{[1cm] gray(0cm)=(1) gray(10cm)=(0)},
		colorbar style={
			yticklabel style={
				font=\footnotesize,
				/pgf/number format/.cd,
				fixed,
				precision=1,
				fixed zerofill,
			},
		},
		enlargelimits=false,
		axis on top,
		point meta min=0,
		point meta max=1,
		xtick=data,
		ytick=data,
		xmin=0,
		xmax=1.8,
		ymin=0,
		ymax=0.42,
		custom_axis_style1,
		custom_axis_style2
	]
		\addplot [matrix plot*,point meta=explicit] file [meta=index 2]
		{make_outputs/phasetran_\algname.txt};
	\end{axis}
\end{tikzpicture}
	\end{minipage}%
	\begin{minipage}[t]{0.32\linewidth}
		\centering
		\renewcommand{\titletext}{VALSE}
		\renewcommand{\algname}{VALSE}
		\pgfplotsset{colorbar=false}
		\subcaption{VALSE}
		\begin{tikzpicture}
	\begin{axis}[
		width=0.951\figurewidth,
		height=\figureheight,
		at={(0\figurewidth,0\figureheight)},
		xlabel=Separation (scaled by $1/N$),
		ylabel=$K/N$,
		colormap={}{[1cm] gray(0cm)=(1) gray(10cm)=(0)},
		colorbar style={
			yticklabel style={
				font=\footnotesize,
				/pgf/number format/.cd,
				fixed,
				precision=1,
				fixed zerofill,
			},
		},
		enlargelimits=false,
		axis on top,
		point meta min=0,
		point meta max=1,
		xtick=data,
		ytick=data,
		xmin=0,
		xmax=1.8,
		ymin=0,
		ymax=0.42,
		custom_axis_style1,
		custom_axis_style2
	]
		\addplot [matrix plot*,point meta=explicit] file [meta=index 2]
		{make_outputs/phasetran_\algname.txt};
	\end{axis}
\end{tikzpicture}
	\end{minipage}%
	\begin{minipage}[t]{0.36\linewidth}
		\centering
		\renewcommand{\titletext}{AST}
		\renewcommand{\algname}{AST}
		\pgfplotsset{colorbar=true}
		\captionsetup[sub]{margin={0mm,0cm}}
		\subcaption{AST}
		\vspace{-5pt}
		\begin{tikzpicture}
	\begin{axis}[
		width=0.951\figurewidth,
		height=\figureheight,
		at={(0\figurewidth,0\figureheight)},
		xlabel=Separation (scaled by $1/N$),
		ylabel=$K/N$,
		colormap={}{[1cm] gray(0cm)=(1) gray(10cm)=(0)},
		colorbar style={
			yticklabel style={
				font=\footnotesize,
				/pgf/number format/.cd,
				fixed,
				precision=1,
				fixed zerofill,
			},
		},
		enlargelimits=false,
		axis on top,
		point meta min=0,
		point meta max=1,
		xtick=data,
		ytick=data,
		xmin=0,
		xmax=1.8,
		ymin=0,
		ymax=0.42,
		custom_axis_style1,
		custom_axis_style2
	]
		\addplot [matrix plot*,point meta=explicit] file [meta=index 2]
		{make_outputs/phasetran_\algname.txt};
	\end{axis}
\end{tikzpicture}
	\end{minipage}
	\\[1mm]
	\begin{minipage}[t]{0.32\linewidth}
		\centering
		\renewcommand{\titletext}{GLS}
		\renewcommand{\algname}{GLS}
		\pgfplotsset{colorbar=false}
		\subcaption{GLS}
		\begin{tikzpicture}
	\begin{axis}[
		width=0.951\figurewidth,
		height=\figureheight,
		at={(0\figurewidth,0\figureheight)},
		xlabel=Separation (scaled by $1/N$),
		ylabel=$K/N$,
		colormap={}{[1cm] gray(0cm)=(1) gray(10cm)=(0)},
		colorbar style={
			yticklabel style={
				font=\footnotesize,
				/pgf/number format/.cd,
				fixed,
				precision=1,
				fixed zerofill,
			},
		},
		enlargelimits=false,
		axis on top,
		point meta min=0,
		point meta max=1,
		xtick=data,
		ytick=data,
		xmin=0,
		xmax=1.8,
		ymin=0,
		ymax=0.42,
		custom_axis_style1,
		custom_axis_style2
	]
		\addplot [matrix plot*,point meta=explicit] file [meta=index 2]
		{make_outputs/phasetran_\algname.txt};
	\end{axis}
\end{tikzpicture}
	\end{minipage}%
	\begin{minipage}[t]{0.32\linewidth}
		\centering
		\renewcommand{\titletext}{ESPRIT}
		\renewcommand{\algname}{ESPRIT}
		\pgfplotsset{colorbar=false}
		\subcaption{ESPRIT}
		\begin{tikzpicture}
	\begin{axis}[
		width=0.951\figurewidth,
		height=\figureheight,
		at={(0\figurewidth,0\figureheight)},
		xlabel=Separation (scaled by $1/N$),
		ylabel=$K/N$,
		colormap={}{[1cm] gray(0cm)=(1) gray(10cm)=(0)},
		colorbar style={
			yticklabel style={
				font=\footnotesize,
				/pgf/number format/.cd,
				fixed,
				precision=1,
				fixed zerofill,
			},
		},
		enlargelimits=false,
		axis on top,
		point meta min=0,
		point meta max=1,
		xtick=data,
		ytick=data,
		xmin=0,
		xmax=1.8,
		ymin=0,
		ymax=0.42,
		custom_axis_style1,
		custom_axis_style2
	]
		\addplot [matrix plot*,point meta=explicit] file [meta=index 2]
		{make_outputs/phasetran_\algname.txt};
	\end{axis}
\end{tikzpicture}
	\end{minipage}%
	\begin{minipage}[t]{0.36\linewidth}
		\centering
		\setlength\figureheight{24.2mm}
		\setlength\figurewidth{55mm}
		\subcaption{Contour at $75\,\%$ success rate}
		\label{contour}
		\vspace{-5pt}
%
\definecolor{mycolor1}{rgb}{1.00000,0.00000,1.00000}%
\begin{tikzpicture}

\begin{axis}[%
width=0.951\figurewidth,
height=\figureheight,
at={(0\figurewidth,0\figureheight)},
scale only axis,
xmin=0.1,
xmax=1.7,
xlabel style={font=\color{white!15!black}},
xlabel={Separation (scaled by $1/N$)},
ymin=0.03,
ymax=0.39,
ylabel style={font=\color{white!15!black}},
ylabel={$K/N$},
axis background/.style={fill=white},
xmajorgrids,
ymajorgrids,
legend style={at={(0.03,0.97)}, anchor=north west, legend cell align=left, align=left, draw=white!15!black},
custom_axis_style1, custom_axis_style2,
xtick = {1.0000e-01,3.0000e-01,5.0000e-01,7.0000e-01,9.0000e-01,1.1000e+00,1.3000e+00,1.5000e+00,1.7000e+00},
ytick = {3.0000e-02,9.0000e-02,1.5000e-01,2.1000e-01,2.7000e-01,3.3000e-01,3.9000e-01}
]
\addplot [color=red, line width=1.0pt, mark=square, mark options={solid, red}]
  table[row sep=crcr]{%
0.29	0.03\\
0.3	0.0322222222222222\\
0.5	0.0710526315789474\\
0.568571428571429	0.09\\
0.688	0.15\\
0.7	0.157826086956522\\
0.817647058823529	0.21\\
0.891666666666667	0.27\\
0.9	0.274\\
1.06969696969697	0.33\\
1.1	0.372857142857143\\
1.11904761904762	0.39\\
};
\addlegendentry{Superfast LSE}

\addplot [color=black, dashed, line width=1.0pt, mark=asterisk, mark options={solid, black}]
  table[row sep=crcr]{%
0.36875	0.03\\
0.5	0.05625\\
0.7	0.0784615384615385\\
0.755555555555556	0.09\\
0.9	0.142\\
0.936363636363637	0.15\\
1.05789473684211	0.21\\
1.1	0.226\\
1.19565217391304	0.27\\
1.3	0.296666666666667\\
1.5	0.30375\\
1.7	0.293076923076923\\
};
\addlegendentry{VALSE}

\addplot [color=mycolor1, dotted, line width=1.0pt, mark=diamond, mark options={solid, mycolor1}]
  table[row sep=crcr]{%
1.7	0.250588235294118\\
1.5	0.247894736842105\\
1.3	0.2625\\
1.20666666666667	0.21\\
1.18	0.15\\
1.3	0.102\\
1.5	0.1\\
1.58571428571429	0.09\\
1.7	0.0814285714285714\\
};
\addlegendentry{AST}

\addplot [color=blue, line width=1.0pt, mark=o, mark options={solid, blue}]
  table[row sep=crcr]{%
0.404	0.03\\
0.5	0.075\\
0.553333333333333	0.09\\
0.684313725490196	0.15\\
0.7	0.161428571428571\\
0.87	0.21\\
0.9	0.222857142857143\\
0.975	0.21\\
1.1	0.182727272727273\\
1.3	0.188181818181818\\
1.38888888888889	0.21\\
1.5	0.2225\\
1.59268292682927	0.27\\
1.7	0.306666666666667\\
};
\addlegendentry{GLS}

\addplot [color=green, dashed, line width=1.0pt, mark=triangle, mark options={solid, rotate=180, green}]
  table[row sep=crcr]{%
0.43953488372093	0.03\\
0.5	0.0421875\\
0.7	0.0827272727272727\\
0.730769230769231	0.09\\
0.9	0.1428\\
0.9375	0.15\\
1.05333333333333	0.21\\
1.1	0.231\\
1.23	0.27\\
1.3	0.29625\\
1.46363636363636	0.33\\
1.5	0.334285714285714\\
1.7	0.361578947368421\\
};
\addlegendentry{ESPRIT}

\end{axis}
\end{tikzpicture}%
	\end{minipage}
	\caption{Simulation results showing phase transitions with complete data
	($\bPhi=\bI$). The plots show block success rate. The set of frequencies are
	generated as closely located pairs, following the methodology described in
	the caption of Fig. \ref{fig:pairsep}. The number of pairs are selected such
	that the specified ratio $K/N$ is achieved as closely as possible. The
	signal length is $N=M=128$ and the SNR is $20\db$.
	Results are averaged over $120$ Monte Carlo trials.}
	\label{fig:phasetran}
\end{figure*}
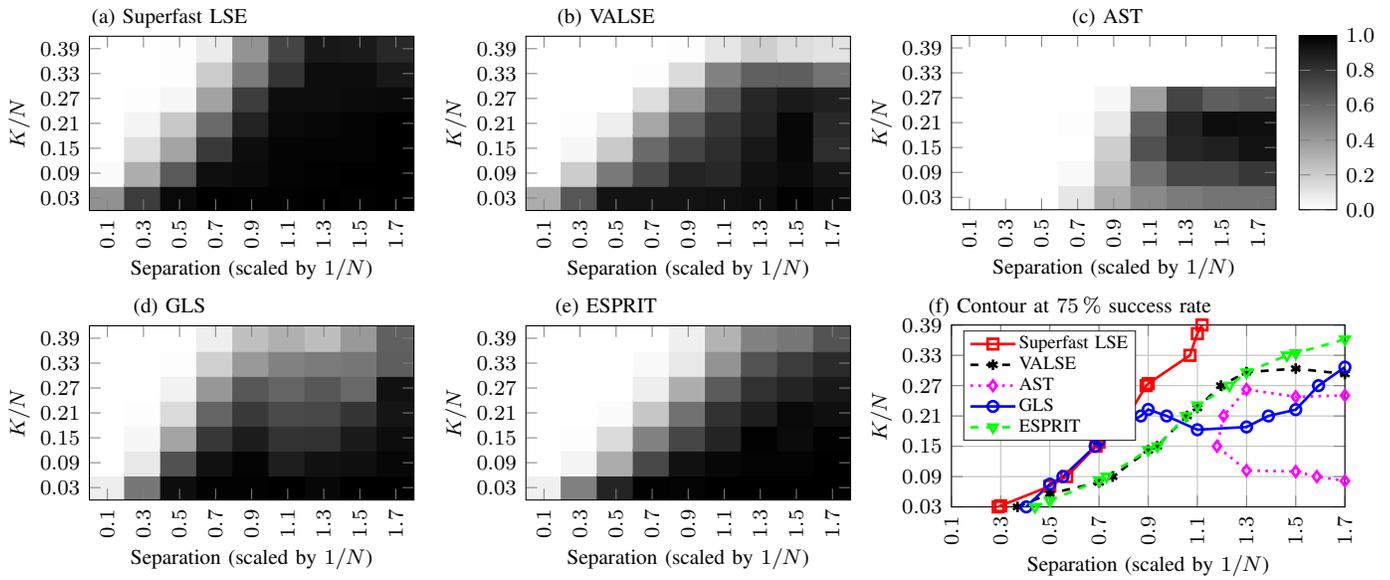

\subsection{Estimation with Incomplete Data}
Fig. \ref{fig:MtoN} reports the performance in the incomplete data case. The
measurement matrix $\bPhi$ is generated by randomly selecting $M$ rows of the
$N\times N$ identity matrix. The set of observation indices is chosen to
include the first and last indices, while the remaining $M-2$ indices are
obtained by uniform random sampling without replacement. Only a subset of the
algorithms are applicable in this case. Our proposed algorithm is implemented
using the techniques described in Sec. \ref{sec:semifast}. We refer to it as
Semifast LSE.

Semifast LSE and VALSE largely show the same performance, while GLS has a
slightly higher NMSE for $M/N\le0.5$. The higher NMSE is caused by a few
outliers (less than $1\%$ of the Monte Carlo trials) where GLS
significantly underestimates the model order. LASSO is again observed to have
reasonable NMSE and CSR while being unable to correctly estimate the set of
frequencies (i.e., BSR$=0$).

\setlength\figureheight{28mm}
\setlength\figurewidth{80mm}
\pgfplotsset{custom_axis_style2/.style={
	log basis x={2},
	legend columns=1,
	legend pos = north east,
	xmax=2^20
}} 
\begin{figure*}[t]
	\begin{minipage}[t]{0.5\linewidth}
		\pgfplotsset{custom_axis_style2/.append style={
			ymin=1e-1,ymax=1e3}}
		\centering
%
\definecolor{mycolor1}{rgb}{0.00000,1.00000,1.00000}%
\definecolor{mycolor2}{rgb}{1.00000,0.00000,1.00000}%
\begin{tikzpicture}

\begin{axis}[%
width=0.951\figurewidth,
height=\figureheight,
at={(0\figurewidth,0\figureheight)},
scale only axis,
xmode=log,
xmin=64,
xmax=32768,
xminorticks=true,
xlabel style={font=\color{white!15!black}},
xlabel={$N$},
ymode=log,
ymin=0.001,
ymax=1000,
yminorticks=true,
ylabel style={font=\color{white!15!black}},
ylabel={Runtime in seconds},
axis background/.style={fill=white},
xmajorgrids,
xminorgrids,
ymajorgrids,
yminorgrids,
legend style={legend cell align=left, align=left, draw=white!15!black},
custom_axis_style1, custom_axis_style2,
xtick = {6.4000e+01,2.5600e+02,1.0240e+03,4.0960e+03,1.6384e+04,6.5536e+04,2.6214e+05,1.0486e+06,4.1943e+06},
ytick = {1.0e-05,1.0e-04,1.0e-03,1.0e-02,1.0e-01,1.0e+00,1.0e+01,1.0e+02,1.0e+03,1.0e+04},
minor ytick = {1.0e-05,2.0e-05,3.0e-05,4.0e-05,5.0e-05,6.0e-05,7.0e-05,8.0e-05,9.0e-05,1.0e-04,2.0e-04,3.0e-04,4.0e-04,5.0e-04,6.0e-04,7.0e-04,8.0e-04,9.0e-04,1.0e-03,2.0e-03,3.0e-03,4.0e-03,5.0e-03,6.0e-03,7.0e-03,8.0e-03,9.0e-03,1.0e-02,2.0e-02,3.0e-02,4.0e-02,5.0e-02,6.0e-02,7.0e-02,8.0e-02,9.0e-02,1.0e-01,2.0e-01,3.0e-01,4.0e-01,5.0e-01,6.0e-01,7.0e-01,8.0e-01,9.0e-01,1.0e+00,2.0e+00,3.0e+00,4.0e+00,5.0e+00,6.0e+00,7.0e+00,8.0e+00,9.0e+00,1.0e+01,2.0e+01,3.0e+01,4.0e+01,5.0e+01,6.0e+01,7.0e+01,8.0e+01,9.0e+01,1.0e+02,2.0e+02,3.0e+02,4.0e+02,5.0e+02,6.0e+02,7.0e+02,8.0e+02,9.0e+02,1.0e+03,2.0e+03,3.0e+03,4.0e+03,5.0e+03,6.0e+03,7.0e+03,8.0e+03,9.0e+03,1.0e+04,2.0e+04,3.0e+04,4.0e+04,5.0e+04,6.0e+04,7.0e+04,8.0e+04,9.0e+04}
]
\addplot [color=red, line width=1.0pt, mark=square, mark options={solid, red}]
  table[row sep=crcr]{%
64	0.104999999999964\\
128	0.254000000000349\\
256	0.233499999999742\\
512	0.323999999999972\\
1024	0.472500000000304\\
2048	0.636999999999796\\
4096	1.00200000000058\\
8192	1.98100000000004\\
16384	5.12600000000022\\
32768	12.4070000000002\\
};
\addlegendentry{Superfast LSE}

\addplot [color=mycolor1, dotted, line width=1.0pt, mark=x, mark options={solid, mycolor1}]
  table[row sep=crcr]{%
64	0.0404999999998751\\
128	0.0889999999996832\\
256	0.201000000000482\\
512	0.539499999999535\\
1024	1.31150000000023\\
2048	3.3069999999998\\
4096	9.65099999999999\\
8192	25.0490000000001\\
16384	69.8934999999998\\
32768	262.4335\\
};
\addlegendentry{LASSO}

\addplot [color=green, dashed, line width=1.0pt, mark=triangle, mark options={solid, rotate=180, green}]
  table[row sep=crcr]{%
64	0.00550000000015558\\
128	0.0145000000002611\\
256	0.0324999999996322\\
512	0.151000000000258\\
1024	1.03700000000001\\
2048	7.42900000000001\\
4096	53.4249999999996\\
8192	404.2425\\
};
\addlegendentry{ESPRIT}

\addplot [color=black, dashed, line width=1.0pt, mark=asterisk, mark options={solid, black}]
  table[row sep=crcr]{%
64	1.08200000000017\\
128	20.483\\
256	40.4585000000003\\
512	65.5515000000005\\
1024	322.725\\
};
\addlegendentry{VALSE}

\addplot [color=mycolor2, dotted, line width=1.0pt, mark=diamond, mark options={solid, mycolor2}]
  table[row sep=crcr]{%
64	1.09299999999984\\
128	2.26300000000073\\
256	14.6349999999998\\
512	134.314\\
};
\addlegendentry{AST}

\addplot [color=blue, line width=1.0pt, mark=o, mark options={solid, blue}]
  table[row sep=crcr]{%
64	1.2250000000001\\
128	6.52500000000007\\
256	49.3965000000007\\
512	466.260499999999\\
};
\addlegendentry{GLS}

\addplot [color=black, line width=1.0pt]
  table[row sep=crcr]{%
2048	0.388874263693869\\
2167.22413992406	0.417643001058536\\
2293.3889026707	0.448491357460874\\
2426.89833598725	0.481566758329722\\
2568.18000922516	0.517026908585781\\
2717.68638264803	0.555040499322872\\
2875.89625645396	0.595787962481924\\
3043.31630415247	0.639462276742586\\
3220.48269520689	0.686269828072131\\
3407.96281213872	0.736431328600115\\
3606.35706759303	0.790182797731294\\
3816.30082718426	0.847776609669368\\
4038.46644428022	0.909482611801318\\
4273.56541324091	0.975589318687637\\
4522.35064800794	1.04640518671872\\
4785.61889334181	1.12225997483336\\
5064.21327642923	1.20350619705321\\
5359.02600703191	1.29052067296818\\
5671.00123482442	1.38370618271461\\
6001.1380730716	1.48349323342069\\
6350.49379832917	1.59034194455567\\
6720.18723641457	1.70474406011114\\
7111.40234549181	1.82722509606685\\
7525.39200774534	1.95834663215248\\
7963.4820417858	2.09870875751175\\
8427.07544863766	2.24895268050985\\
8917.65690490671	2.40976351360105\\
9436.79751751676	2.58187324489354\\
9986.15985524307	2.76606390881605\\
10567.5032731559	2.96317096910843\\
11182.6895470264	3.17408692822952\\
11833.6888357383	3.39976517820388\\
12522.5859908018	3.64122410891796\\
13251.5872331741	3.89955149092937\\
14023.0272187716	4.1759091509761\\
14839.3765152999	4.47153795956743\\
15703.2495143457	4.78776315131261\\
16617.4128040727	5.12600000000022\\
17584.7940293308	5.48775987188651\\
18608.4912675578	5.874656682191\\
19691.7829504967	6.2884137814378\\
20838.1383635066	6.73087130002784\\
22051.2287560885	7.20399398128916\\
23334.9390992103	7.70987953523319\\
24693.3805270832	8.25076754735759\\
26130.9035032351	8.82904897908398\\
27652.1117530456	9.44727629881411\\
29261.8770073625	10.1081742851393\\
30965.3546044165	10.8146515464535\\
32768	11.5698128041151\\
};
\addlegendentry{$\calo(N\log^2\!N)$}

\end{axis}
\end{tikzpicture}%
	\end{minipage}%
	\begin{minipage}[t]{0.5\linewidth}
		\pgfplotsset{custom_axis_style2/.append style={
			ymin=1e-1,ymax=1e3}}
		\centering
%
\definecolor{mycolor1}{rgb}{0.00000,1.00000,1.00000}%
\begin{tikzpicture}

\begin{axis}[%
width=0.951\figurewidth,
height=\figureheight,
at={(0\figurewidth,0\figureheight)},
scale only axis,
xmode=log,
xmin=64,
xmax=32768,
xminorticks=true,
xlabel style={font=\color{white!15!black}},
xlabel={$N$},
ymode=log,
ymin=0.01,
ymax=1000,
yminorticks=true,
ylabel style={font=\color{white!15!black}},
ylabel={Runtime in seconds},
axis background/.style={fill=white},
xmajorgrids,
xminorgrids,
ymajorgrids,
yminorgrids,
legend style={legend cell align=left, align=left, draw=white!15!black},
custom_axis_style1, custom_axis_style2,
xtick = {6.4000e+01,2.5600e+02,1.0240e+03,4.0960e+03,1.6384e+04,6.5536e+04,2.6214e+05,1.0486e+06,4.1943e+06},
ytick = {1.0e-05,1.0e-04,1.0e-03,1.0e-02,1.0e-01,1.0e+00,1.0e+01,1.0e+02,1.0e+03,1.0e+04},
minor ytick = {1.0e-05,2.0e-05,3.0e-05,4.0e-05,5.0e-05,6.0e-05,7.0e-05,8.0e-05,9.0e-05,1.0e-04,2.0e-04,3.0e-04,4.0e-04,5.0e-04,6.0e-04,7.0e-04,8.0e-04,9.0e-04,1.0e-03,2.0e-03,3.0e-03,4.0e-03,5.0e-03,6.0e-03,7.0e-03,8.0e-03,9.0e-03,1.0e-02,2.0e-02,3.0e-02,4.0e-02,5.0e-02,6.0e-02,7.0e-02,8.0e-02,9.0e-02,1.0e-01,2.0e-01,3.0e-01,4.0e-01,5.0e-01,6.0e-01,7.0e-01,8.0e-01,9.0e-01,1.0e+00,2.0e+00,3.0e+00,4.0e+00,5.0e+00,6.0e+00,7.0e+00,8.0e+00,9.0e+00,1.0e+01,2.0e+01,3.0e+01,4.0e+01,5.0e+01,6.0e+01,7.0e+01,8.0e+01,9.0e+01,1.0e+02,2.0e+02,3.0e+02,4.0e+02,5.0e+02,6.0e+02,7.0e+02,8.0e+02,9.0e+02,1.0e+03,2.0e+03,3.0e+03,4.0e+03,5.0e+03,6.0e+03,7.0e+03,8.0e+03,9.0e+03,1.0e+04,2.0e+04,3.0e+04,4.0e+04,5.0e+04,6.0e+04,7.0e+04,8.0e+04,9.0e+04}
]
\addplot [color=red, line width=1.0pt, mark=square, mark options={solid, red}]
  table[row sep=crcr]{%
64	0.149999999999991\\
128	0.238500000000363\\
256	0.24399999999987\\
512	0.351499999999787\\
1024	0.497999999999843\\
2048	0.699499999999443\\
4096	0.930000000000018\\
8192	1.83400000000006\\
16384	4.10349999999978\\
32768	9.42650000000019\\
};
\addlegendentry{Semifast LSE}

\addplot [color=mycolor1, dotted, line width=1.0pt, mark=x, mark options={solid, mycolor1}]
  table[row sep=crcr]{%
64	0.0484999999998237\\
128	0.103500000000011\\
256	0.213499999999996\\
512	0.545500000000064\\
1024	1.32400000000006\\
2048	3.28550000000023\\
4096	9.6180000000001\\
8192	23.5460000000002\\
16384	62.4345\\
32768	244.7985\\
};
\addlegendentry{LASSO}

\addplot [color=black, dashed, line width=1.0pt, mark=asterisk, mark options={solid, black}]
  table[row sep=crcr]{%
64	1.1980000000001\\
128	4.33399999999996\\
256	62.131\\
512	56.808\\
1024	438.6515\\
};
\addlegendentry{VALSE}

\addplot [color=blue, line width=1.0pt, mark=o, mark options={solid, blue}]
  table[row sep=crcr]{%
64	1.18799999999991\\
128	6.28099999999985\\
256	48.6239999999999\\
512	447.639\\
};
\addlegendentry{GLS}

\addplot [color=black, line width=1.0pt]
  table[row sep=crcr]{%
2048	0.39678295694816\\
2167.22413992406	0.42299764666231\\
2293.3889026707	0.450919820176502\\
2426.89833598725	0.480659438575343\\
2568.18000922516	0.512333479071585\\
2717.68638264803	0.546066379237214\\
2875.89625645396	0.581990509178805\\
3043.31630415247	0.62024667340518\\
3220.48269520689	0.660984644244269\\
3407.96281213872	0.704363728781612\\
3606.35706759303	0.750553371415684\\
3816.30082718426	0.799733794255576\\
4038.46644428022	0.852096677724944\\
4273.56541324091	0.907845883883162\\
4522.35064800794	0.967198225130672\\
4785.61889334181	1.03038428113129\\
5064.21327642923	1.09764926696025\\
5359.02600703191	1.16925395567362\\
5671.00123482442	1.2454756586933\\
6001.1380730716	1.3266092676124\\
6350.49379832917	1.41296836124973\\
6720.18723641457	1.50488638201969\\
7111.40234549181	1.60271788593613\\
7525.39200774534	1.70683987083665\\
7963.4820417858	1.81765318769841\\
8427.07544863766	1.93558404021835\\
8917.65690490671	2.06108557815154\\
9436.79751751676	2.19463959024191\\
9986.15985524307	2.33675830294106\\
10567.5032731559	2.48798629149468\\
11182.6895470264	2.6489025103839\\
11833.6888357383	2.82012245054107\\
12522.5859908018	3.00230043121966\\
13251.5872331741	3.19613203488473\\
14023.0272187716	3.40235669400927\\
14839.3765152999	3.62176043921059\\
15703.2495143457	3.85517881874515\\
16617.4128040727	4.10349999999978\\
17584.7940293308	4.36766806427469\\
18608.4912675578	4.64868650685268\\
19691.7829504967	4.94762195508985\\
20838.1383635066	5.26560811805031\\
22051.2287560885	5.60384998204324\\
23334.9390992103	5.9636282673071\\
24693.3805270832	6.34630416202809\\
26130.9035032351	6.75332435087904\\
27652.1117530456	7.18622635632616\\
29261.8770073625	7.64664421207785\\
30965.3546044165	8.13631448924479\\
32768	8.65708269705076\\
};
\addlegendentry{$\calo(N\log N)$}

\end{axis}
\end{tikzpicture}%
	\end{minipage}%
	\vspace{-3mm}
	\caption{
		Runtimes in seconds versus problem size $N$. We show for both complete
		(left) and incomplete (right) data case. The number of components is
		$K=15$ and the SNR is $20\db$. Values are averaged over 20 Monte Carlo
		trials. In the incomplete data case we set $M=0.75N$. We also plot (solid black) the
		asymptotic per-iteration complexity of Superfast and Semifast LSE.}
	\label{fig:runtime_N}
\end{figure*}
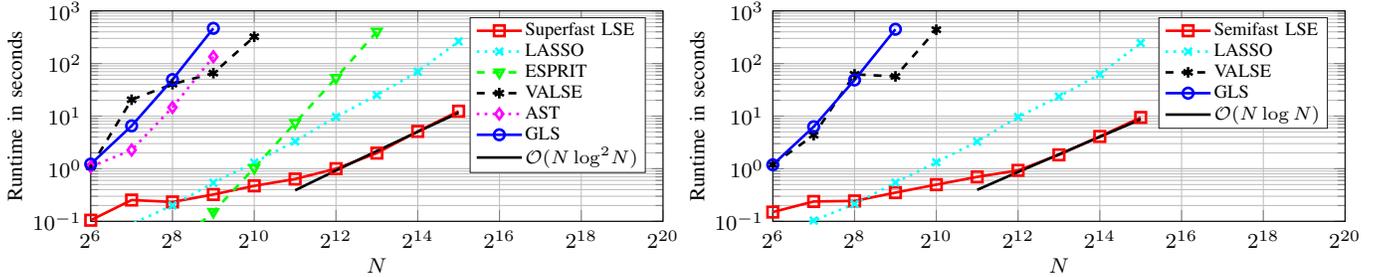

\pgfplotsset{custom_axis_style2/.style={
	legend columns=1,
	legend pos=north west,
	xticklabel={
        \pgfkeys{/pgf/fpu=true}
		\pgfmathparse{exp(\tick)}%
		\pgfmathprintnumber[fixed, precision=0]{\pgfmathresult}
        \pgfkeys{/pgf/fpu=false}
      }
}}
\begin{figure*}[t]
	\begin{minipage}[t]{0.5\linewidth}
		\pgfplotsset{custom_axis_style2/.append style={
			ymin=1e-1,ymax=1e3}}
		\centering
%
\definecolor{mycolor1}{rgb}{0.00000,1.00000,1.00000}%
\begin{tikzpicture}

\begin{axis}[%
width=0.951\figurewidth,
height=\figureheight,
at={(0\figurewidth,0\figureheight)},
scale only axis,
xmode=log,
xmin=4,
xmax=1024,
xminorticks=true,
xlabel style={font=\color{white!15!black}},
xlabel={$K$},
ymode=log,
ymin=0.647500000000582,
ymax=9.95599999999977,
yminorticks=true,
ylabel style={font=\color{white!15!black}},
ylabel={Runtime in seconds},
axis background/.style={fill=white},
xmajorgrids,
xminorgrids,
ymajorgrids,
yminorgrids,
legend style={legend cell align=left, align=left, draw=white!15!black},
custom_axis_style1, custom_axis_style2,
xtick = {4.0000e+00,8.0000e+00,1.6000e+01,3.2000e+01,6.4000e+01,1.2800e+02,2.5600e+02,5.1200e+02,1.0240e+03},
ytick = {1.0e-05,1.0e-04,1.0e-03,1.0e-02,1.0e-01,1.0e+00,1.0e+01,1.0e+02,1.0e+03,1.0e+04},
minor ytick = {1.0e-05,2.0e-05,3.0e-05,4.0e-05,5.0e-05,6.0e-05,7.0e-05,8.0e-05,9.0e-05,1.0e-04,2.0e-04,3.0e-04,4.0e-04,5.0e-04,6.0e-04,7.0e-04,8.0e-04,9.0e-04,1.0e-03,2.0e-03,3.0e-03,4.0e-03,5.0e-03,6.0e-03,7.0e-03,8.0e-03,9.0e-03,1.0e-02,2.0e-02,3.0e-02,4.0e-02,5.0e-02,6.0e-02,7.0e-02,8.0e-02,9.0e-02,1.0e-01,2.0e-01,3.0e-01,4.0e-01,5.0e-01,6.0e-01,7.0e-01,8.0e-01,9.0e-01,1.0e+00,2.0e+00,3.0e+00,4.0e+00,5.0e+00,6.0e+00,7.0e+00,8.0e+00,9.0e+00,1.0e+01,2.0e+01,3.0e+01,4.0e+01,5.0e+01,6.0e+01,7.0e+01,8.0e+01,9.0e+01,1.0e+02,2.0e+02,3.0e+02,4.0e+02,5.0e+02,6.0e+02,7.0e+02,8.0e+02,9.0e+02,1.0e+03,2.0e+03,3.0e+03,4.0e+03,5.0e+03,6.0e+03,7.0e+03,8.0e+03,9.0e+03,1.0e+04,2.0e+04,3.0e+04,4.0e+04,5.0e+04,6.0e+04,7.0e+04,8.0e+04,9.0e+04}
]
\addplot [color=red, line width=1.0pt, mark=square, mark options={solid, red}]
  table[row sep=crcr]{%
4	0.647500000000582\\
8	0.722500000000582\\
16	0.905499999998938\\
32	1.07400000000052\\
64	1.40049999999937\\
128	1.74999999999964\\
256	2.23750000000073\\
512	3.87200000000012\\
1024	8.71749999999993\\
};
\addlegendentry{Superfast LSE}

\addplot [color=mycolor1, dotted, line width=1.0pt, mark=x, mark options={solid, mycolor1}]
  table[row sep=crcr]{%
4	9.59099999999962\\
8	9.95599999999977\\
16	9.53700000000026\\
32	9.9255000000001\\
64	9.45849999999991\\
128	8.46949999999997\\
256	7.43350000000064\\
512	7.05000000000036\\
1024	7.85399999999972\\
};
\addlegendentry{LASSO}

\addplot [color=black, line width=1.0pt]
  table[row sep=crcr]{%
128	0.954403202652189\\
133.548919163778	0.995777470005065\\
139.338389139167	1.03894535036577\\
145.378838027795	1.08398459852798\\
151.681145999203	1.13097634002808\\
158.256664888413	1.18000521726997\\
165.117238643061	1.23115954198456\\
172.275224656942	1.28453145429852\\
179.743516028372	1.34021708869878\\
187.535564783488	1.39831674719178\\
195.665406106277	1.45893507996932\\
204.147683619024	1.52218127390643\\
212.99767575867	1.58816924923083\\
222.231323296619	1.65701786471818\\
231.86525805156	1.7288511317827\\
241.916832847003	1.80379843784886\\
252.404152767514	1.88199477940646\\
263.346107769927	1.9635810051688\\
274.762406708294	2.04870406977203\\
286.673612833837	2.13751729847256\\
299.101180833863	2.23018066331946\\
312.067495476346	2.326861071299\\
325.595911929791	2.42773266497073\\
339.710797831003	2.53297713613625\\
354.437577176526	2.64278405310593\\
369.802776116823	2.75735120215286\\
385.834070735675	2.87688494376922\\
402.560336900861	3.00160058436661\\
420.011702275911	3.13172276409005\\
438.219600586614	3.26748586144396\\
457.216828240039	3.40913441545913\\
477.037603398032	3.55692356616099\\
497.717627611612	3.7111195141326\\
519.294150127272	3.87200000000012\\
541.806034981016	4.03985480470442\\
565.293831000992	4.21498627145994\\
589.799844844798	4.39770985034093\\
615.368217203023	4.58835466647604\\
642.045002306289	4.78726411287464\\
669.878250878988	4.99479646895276\\
698.91809668915	5.21132554587264\\
729.216846850322	5.43724135985843\\
760.829076038118	5.67295083470067\\
793.811724791147	5.91887853471507\\
828.224202073378	6.17546742947567\\
864.128492282676	6.44317969169995\\
901.589266898256	6.72249752972293\\
940.674000968167	7.0139240560599\\
981.453094646609	7.31798419362207\\
1024	7.63522562121751\\
};
\addlegendentry{$\calo(K)$}

\end{axis}
\end{tikzpicture}%
	\end{minipage}%
	\begin{minipage}[t]{0.5\linewidth}
		\pgfplotsset{custom_axis_style2/.append style={
			ymin=1e-1,ymax=1e3}}
		\centering
%
\definecolor{mycolor1}{rgb}{0.00000,1.00000,1.00000}%
\begin{tikzpicture}

\begin{axis}[%
width=0.951\figurewidth,
height=\figureheight,
at={(0\figurewidth,0\figureheight)},
scale only axis,
xmode=log,
xmin=4,
xmax=1024,
xminorticks=true,
xlabel style={font=\color{white!15!black}},
xlabel={$K$},
ymode=log,
ymin=0.1,
ymax=1000,
yminorticks=true,
ylabel style={font=\color{white!15!black}},
ylabel={Runtime in seconds},
axis background/.style={fill=white},
xmajorgrids,
xminorgrids,
ymajorgrids,
yminorgrids,
legend style={legend cell align=left, align=left, draw=white!15!black},
custom_axis_style1, custom_axis_style2,
xtick = {4.0000e+00,8.0000e+00,1.6000e+01,3.2000e+01,6.4000e+01,1.2800e+02,2.5600e+02,5.1200e+02,1.0240e+03},
ytick = {1.0e-05,1.0e-04,1.0e-03,1.0e-02,1.0e-01,1.0e+00,1.0e+01,1.0e+02,1.0e+03,1.0e+04},
minor ytick = {1.0e-05,2.0e-05,3.0e-05,4.0e-05,5.0e-05,6.0e-05,7.0e-05,8.0e-05,9.0e-05,1.0e-04,2.0e-04,3.0e-04,4.0e-04,5.0e-04,6.0e-04,7.0e-04,8.0e-04,9.0e-04,1.0e-03,2.0e-03,3.0e-03,4.0e-03,5.0e-03,6.0e-03,7.0e-03,8.0e-03,9.0e-03,1.0e-02,2.0e-02,3.0e-02,4.0e-02,5.0e-02,6.0e-02,7.0e-02,8.0e-02,9.0e-02,1.0e-01,2.0e-01,3.0e-01,4.0e-01,5.0e-01,6.0e-01,7.0e-01,8.0e-01,9.0e-01,1.0e+00,2.0e+00,3.0e+00,4.0e+00,5.0e+00,6.0e+00,7.0e+00,8.0e+00,9.0e+00,1.0e+01,2.0e+01,3.0e+01,4.0e+01,5.0e+01,6.0e+01,7.0e+01,8.0e+01,9.0e+01,1.0e+02,2.0e+02,3.0e+02,4.0e+02,5.0e+02,6.0e+02,7.0e+02,8.0e+02,9.0e+02,1.0e+03,2.0e+03,3.0e+03,4.0e+03,5.0e+03,6.0e+03,7.0e+03,8.0e+03,9.0e+03,1.0e+04,2.0e+04,3.0e+04,4.0e+04,5.0e+04,6.0e+04,7.0e+04,8.0e+04,9.0e+04}
]
\addplot [color=red, line width=1.0pt, mark=square, mark options={solid, red}]
  table[row sep=crcr]{%
4	0.341500000000007\\
8	0.525500000000068\\
16	1.05350000000005\\
32	1.93999999999987\\
64	4.26549999999984\\
128	9.88899999999991\\
256	27.5905000000001\\
512	101.9825\\
1024	583.251\\
};
\addlegendentry{Semifast LSE}

\addplot [color=mycolor1, dotted, line width=1.0pt, mark=x, mark options={solid, mycolor1}]
  table[row sep=crcr]{%
4	10.4754999999999\\
8	9.79900000000004\\
16	9.9845\\
32	8.9105000000001\\
64	8.25300000000013\\
128	7.60200000000011\\
256	7.96650000000001\\
512	8.39400000000007\\
1024	16.5095000000002\\
};
\addlegendentry{LASSO}

\addplot [color=black, line width=1.0pt]
  table[row sep=crcr]{%
128	1.52726806983639\\
133.548919163778	1.73462857304505\\
139.338389139167	1.97014286218049\\
145.378838027795	2.23763343791057\\
151.681145999203	2.54144179012172\\
158.256664888413	2.8864988622123\\
165.117238643061	3.27840508247638\\
172.275224656942	3.72352126152214\\
179.743516028372	4.22907183103029\\
187.535564783488	4.80326209946301\\
195.665406106277	5.45541142783499\\
204.147683619024	6.19610448704847\\
212.99767575867	7.03736305176496\\
222.231323296619	7.99284111910411\\
231.86525805156	9.07804651903231\\
241.916832847003	10.3105926132749\\
252.404152767514	11.7104841679365\\
263.346107769927	13.300442039668\\
274.762406708294	15.1062719451794\\
286.673612833837	17.1572832993911\\
299.101180833863	19.4867649201498\\
312.067495476346	22.1325253204076\\
325.595911929791	25.1375063570439\\
339.710797831003	28.5504801961199\\
354.437577176526	32.4268409066209\\
369.802776116823	36.8295035305994\\
385.834070735675	41.8299252220244\\
402.560336900861	47.5092650278707\\
420.011702275911	53.9597011352064\\
438.219600586614	61.2859269637764\\
457.216828240039	69.6068503863293\\
477.037603398032	79.0575236557743\\
497.717627611612	89.7913353627454\\
519.294150127272	101.9825\\
541.806034981016	115.828885540387\\
565.293831000992	131.555224921219\\
589.799844844798	149.416763558844\\
615.368217203023	169.703402094207\\
642.045002306289	192.744401607965\\
669.878250878988	218.913727672879\\
698.91809668915	248.636119979815\\
729.216846850322	282.393986050039\\
760.829076038118	320.735230921812\\
793.811724791147	364.282149890544\\
828.224202073378	413.741528635581\\
864.128492282676	469.91611466316\\
901.589266898256	533.71764625208\\
940.674000968167	606.181650367634\\
981.453094646609	688.484249720452\\
1024	781.96125175623\\
};
\addlegendentry{$\calo(K^3)$}

\end{axis}
\end{tikzpicture}%
	\end{minipage}%
	\vspace{-3mm}
	\caption{
		Runtimes in seconds versus number of components $K$. We show for both
		complete (left) and incomplete (right) data case. The problem size is
		$N=4096$ and the SNR is $20\db$.
		Values are averaged over 20 Monte
		Carlo trials. We only show results for algorithms which has runtime
		lower than $10$ s at $K=4$.
		In the incomplete data case we set $M=0.75N=3072$. We also plot (solid
		black) the asymptotic per-iteration complexity of Superfast and
		Semifast LSE.}
	\label{fig:runtime_K}
\end{figure*}
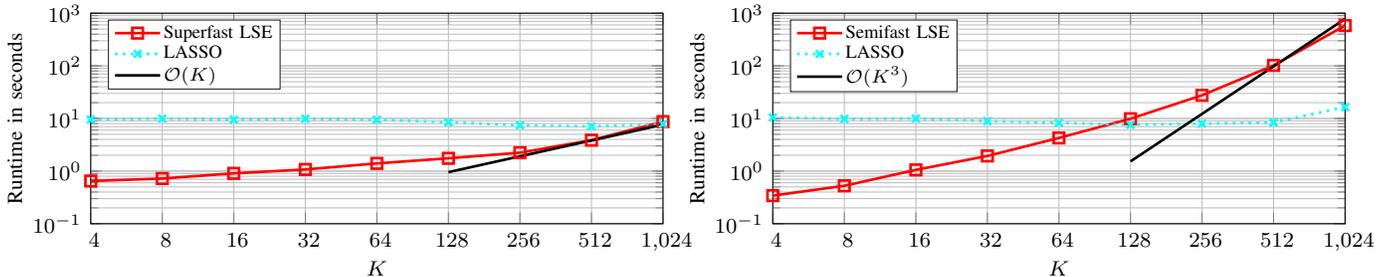

\subsection{Phase Transitions}
Inspired by the concept of phase transitions in compressed sensing, we perform
an experiment which shows a similar phenomenon for LSE. In particular we
demonstrate that for each algorithm there is a region in the space of system
parameters where it can almost perfectly recover the frequencies and a region
where it cannot, with a fairly sharp transition between the two. The results,
in terms of BSR, are seen in Fig. \ref{fig:phasetran}.

We first note that AST has rather poor performance, which is consistent with
the observation in Fig. \ref{fig:snr} that its BSR is significantly below 1.
Turning our attention to VALSE, GLS and ESPRIT, we see that these algorithms
generally do not deal well with a large number of components, in the sense that
the BSR is significantly below 1 for $K/N\ge0.15$. It is seen in Fig.
\ref{contour} that Superfast LSE has the largest region with high probability
of successful recovery ($\mathrm{BSR}\ge0.75$).

\subsection{Computation Times}
In Fig. \ref{fig:runtime_N} and \ref{fig:runtime_K} we show algorithm
runtimes for varying problem sizes. Our proposed method uses the
superfast and semifast implementations in Secs. \ref{sec:superfast} and
\ref{sec:semifast}, respectively. The results are obtained using MATLAB on a 2011
MacBook Pro. To avoid differences in results originating from the amount of parallelism
achieved in each implementation, MATLAB is restricted to only use a single
computational thread. The part of the code where each algorithm spends
significant time is implemented as native code via MATLAB's codegen feature.

For varying $N$ (Fig. \ref{fig:runtime_N}), we first observe that for small to
moderate problem sizes ($N\le2^{10}$) the difference between LASSO and our
proposed algorithms is small (less than 1 second).
This difference is mainly due to implementation details.
In the large-$N$ region, Superfast and Semifast LSE are
approximately an order of magnitude faster than LASSO. We observe that the asymptotic
per-iteration complexity of Superfast and Semifast LSE describes the scaling
of the total runtime well for $N\ge2^{12}$, because the number of iterations
(not shown) stays practically constant.
The state-of-the-art LSE methods VALSE, AST and GLS all
have $\calo(N^3)$ time complexity or worse, which results in very large
runtimes even when the problem size is moderate (e.g. $>100\,\mathrm{s}$ for
AST and GLS at $N=512$). For large problem sizes, the $\calo(N^3)$ time
complexity of ESPRIT is evident and Superfast/Semifast LSE and LASSO significantly
outperform ESPRIT.

In Fig. \ref{fig:runtime_K} we show results illustrating how the
computation time scales with $K$. In this analysis we assume $\hat K=\calo(K)$.
First we note that the runtime of LASSO is practically constant with $K$. In
the complete data case, the per-iteration complexity of Superfast LSE scales
linearly with $K$. In practice we see a slower scaling with $K$,
which means that the values of $K$ we use here are not large enough to reach the
asymptotic region. Simulations with large $K$ cannot be run, because we need
approximately $K<N/4$ for $\hat K=\calo(K)$ to hold (cf. Fig. \ref{fig:phasetran}).

In the incomplete data case, we see that the runtime of Semifast LSE increases
quickly with $K$, such that for $K>128$ LASSO is faster than Semifast
LSE. We do, however, see that the asymptotic complexity of $\calo(K^3)$ is not
reached in our experiment, because the runtime is dominated by the calculation
of $\bs^\calg$, which has complexity $\calo(LK^2)$.

\section{Conclusions}
\label{sec:conclusions}
We have presented a low-complexity algorithm for line spectral estimation.
Computational methods for both the complete and incomplete data cases have been
presented, along with an extension to the case of multiple measurement vectors.

The proposed algorithm falls in the category of Bayesian methods for line spectral
estimation. Bayesian methods are widely accepted due to their high estimation
accuracy, but a drawback of this class of methods has historically been
their large computational complexity. In that respect, this work makes an
important contribution in making Bayesian methods more viable in practice.

At the core of the computational method for the complete data case lies the
application of the Gohberg-Semencul formula to the Toeplitz signal covariance
matrix. Many methods for line spectral estimation have Toeplitz covariance
matrices at their core and we conjecture that the computational complexity of
some of them can be drastically reduced by applying the techniques
we have demonstrated in this paper.

Our numerical experiments show that our Superfast LSE algorithm has very high
estimation accuracy. For example, in Fig. \ref{fig:phasetran} we see that
Superfast LSE attains high frequency recovery rates for a much larger set of
scenarios than any of the reference algorithms. At the same time our algorithm
has so low computation time that it makes highly-accurate LSE feasible for
problems with size much larger than methods currently available in the
literature can practically deal with.

\appendices

\section{Convergence Analysis}
\label{app:convergence}
We now discuss the convergence of our proposed block-coordinate descent
algorithm. To do so we introduce an iteration index $i$ on all estimated
variables. Our algorithm then produces sequences of blocks of estimates denoted as
$\{\hat\bz^i\}$, $\{\hat\zeta^i\}$, $\{\hat\beta^i\}$ and
$\{(\hat\btheta^i, \hat\bgamma^i)\}$.
We denote the value of the objective function at the end of the $i$th
iteration as $\call^i \triangleq \call(\hat\bz^i, \hat\zeta^i, \hat\beta^i,
\hat\btheta^i, \hat\bgamma^i)$.
We first note that all updates of the estimates are guaranteed not to increase
the objective, so $\{\call^i\}$ is a non-increasing sequence. Since
$\beta\ge\varepsilon_\beta$ it is also bounded below and thus it converges.
Therefore, our proposed algorithm terminates in finite time.

Unfortunately, convergence of $\{\call^i\}$ does not imply convergence of the
sequences of estimates. Even with exact minimization in each
block of variables, block-coordinate descent on non-convex functions can get stuck in
an infinite cycle \cite{powell-search}. This is further
complicated by the fact that our algorithm only approximately solves the
minimization in some of the blocks.

Proposition 5 in \cite{grippo-convergence} shows that with exact minimization
in each block, block-coordinate descent converges if the objective
function is strictly quasiconvex in all but 2 blocks. The objective function
$\call$ is strictly quasiconvex in $\zeta$ and $\beta$. There is thus hope that
we can prove convergence of our scheme which, in lieu of computing the exact
minimizer, merely has a descent property in each block. Our approach to show
convergence shares the same overall idea as that in \cite{grippo-convergence},
while many details differ.

To discuss the convergence properties, we first derive a number of lemmas.
Theorem \ref{thm:convergence} is proved at the end of the appendix.
For notational simplicity we take the convention that for each $i$ the
block-coordinate descent algorithm cycles through the block updates in the
following order: $\hat\bz^i, \hat\zeta^i, \hat\beta^i$ and finally
$(\hat\btheta^i, \hat\bgamma^i)$, such that for example $\hat\beta^i$ is found
based on $(\hat\bz^i, \hat\zeta^i, \hat\beta^{i-1}, \hat\btheta^{i-1},
\hat\bgamma^{i-1})$. This is strictly speaking not how we have defined our
algorithm, but that does not affect the correctness of our analysis.

\begin{lemma}
	\label{lem:limitpoints}
	The sequence of estimates has at least one convergent subsequence, i.e.,
	at least one limit point.
\end{lemma}
\begin{IEEEproof}
	All variables but $\bgamma$ and $\beta$ are defined to be in a closed and
	bounded set. Since $\lim_{\beta\rightarrow\infty}\call=\infty$
	we can restrict $\beta$ to a closed and bounded set determined by the (finite)
	initial value of the objective function. A similar argument holds for each
	$\gamma_k$. The lemma then follows from the Bolzano-Weierstrass theorem.
\end{IEEEproof}

\begin{lemma}
	\label{lem:z}
	The sequence $\{\hat\bz^i\}$ converges.
\end{lemma}
\begin{IEEEproof}
	Each activation of a component gives a decrease in $\call$ of at least
	$\varepsilon_\call$. Since $\{\call^i\}$ is lower bounded, there can only be finitely
	many activations. Since there cannot be more deactivations than
	activations, also the number of deactivations is finite. There is thus a
	finite number of changes to $\hat\bz$ and $\{\hat\bz^i\}$ converges. We denote
	the limit point as $\bar\bz$.
\end{IEEEproof}

\begin{lemma}
	\label{lem:zeta}
	The sequence $\{\hat\zeta^i\}$ converges. Further, the limit point $\bar\zeta$ is
	the unique global minimizer of $\zeta\mapsto\call(\bar\bz, \zeta, \beta, \btheta,
	\bgamma)$ for any $\beta$, $\btheta$ and $\bgamma$.
\end{lemma}
\begin{IEEEproof}
	The first statement follows from Lemma \ref{lem:z} since $\hat\zeta^i$
	\eqref{zetaUpdate} is only a function of $\hat\bz^i$. The second statement
	results from the fact that $\hat\zeta^i$ is defined as the global minimizer
	of $\zeta\mapsto\call(\hat\bz^i, \zeta, \hat\beta^{i-1},
	\hat\btheta^{i-1}, \hat\bgamma^{i-1})$ and this global minimizer does not
	depend on $\hat\beta^{i-1}$, $\hat\btheta^{i-1}$ and $\hat\bgamma^{i-1}$.
\end{IEEEproof}

\begin{lemma}
	\label{lem:beta}
	The sequence $\{\hat\beta^i\}$ converges to the limit point $\bar\beta$.
	Further, for every limit point $(\bar\bz, \bar\zeta, \bar\btheta,
	\bar\bgamma)$ of the remaining variables, the limit point $\bar\beta$ is a
	local minimum at the boundary $\varepsilon_\beta$ or a stationary point of
	$\beta\mapsto\call(\bar\bz, \bar\zeta, \beta, \bar\btheta,
	\bar\bgamma)$.
\end{lemma}
\begin{IEEEproof}
	To perform this proof we expand our previous notation and denote the
	upper bound \eqref{bound_explicit} as
		$Q(\beta; \hat\bz^{i}, \hat\beta^{i-1}, \hat\btheta^{i-1},
		\hat\bgamma^{i-1})$. Then,
		\begin{align*}
			\call(\hat\bz^{i}, \hat\zeta^{i}, \hat\beta^{i-1}, \hat\btheta^{i-1},
				\hat\bgamma^{i-1})
			= Q(\hat\beta^{i-1}; \hat\bz^{i}, \hat\beta^{i-1},
				\hat\btheta^{i-1}, \hat\bgamma^{i-1}) \\
			\ge Q(\hat\beta^{i}; \hat\bz^{i}, \hat\beta^{i-1},
				\hat\btheta^{i-1}, \hat\bgamma^{i-1})
			\ge \call(\hat\bz^{i}, \hat\zeta^{i}, \hat\beta^i, \hat\btheta^{i-1},
				\hat\bgamma^{i-1}).
		\end{align*}
		Recalling that $\{\call^i\}$ converges, we have
		\begin{multline*}
			\limm{i\rightarrow\infty}
			\Big|
			\call(\hat\bz^{i}, \hat\zeta^{i}, \hat\beta^{i-1}, \hat\btheta^{i-1},
				\hat\bgamma^{i-1}) \\
			- \call(\hat\bz^{i}, \hat\zeta^{i}, \hat\beta^i, \hat\btheta^{i-1},
				\hat\bgamma^{i-1}) \Big| = 0,
		\end{multline*}
		and thus
		\begin{multline*}
			\limm{i\rightarrow\infty}
			\Big|
			Q(\hat\beta^{i-1}; \hat\bz^{i}, \hat\beta^{i-1},
				\hat\btheta^{i-1}, \hat\bgamma^{i-1}) \\
			- Q(\hat\beta^{i}; \hat\bz^{i}, \hat\beta^{i-1},
				\hat\btheta^{i-1}, \hat\bgamma^{i-1})
				\Big| = 0.
				\shownumber\label{lim_Q}
		\end{multline*}
		Reasoning by contradiction, assume that the sequence of estimates
		$\{\hat\beta^i\}$ has two
		limit points $\bar\beta_1$ and $\bar\beta_2$, such that
		$\bar\beta_1\ne\bar\beta_2$.
		Let $(\bar\btheta, \bar\bgamma)$ be any limit point of
		$\{\hat\btheta^i, \hat\bgamma^i\}$ (such a limit point exists due to
		Lemma \ref{lem:limitpoints}).
		Then by \eqref{lim_Q} we must have
		\begin{align}
			Q(\bar\beta_1; \bar\bz,\bar\beta_1,\bar\btheta,\bar\bgamma)
			=
			Q(\bar\beta_2; \bar\bz,\bar\beta_1,\bar\btheta,\bar\bgamma).
			\label{Q_equal}
		\end{align}
		Recalling the definition of $\hat\beta^{i}$, we have that $\bar\beta_2$
		uniquely minimizes $Q$. Then, since we assumed
		$\bar\beta_1\ne\bar\beta_2$, we get
		\begin{align*}
			Q(\bar\beta_1; \bar\bz,\bar\beta_1,\bar\btheta,\bar\bgamma)
			>
			Q(\bar\beta_2; \bar\bz,\bar\beta_1,\bar\btheta,\bar\bgamma),
		\end{align*}
		which contradicts \eqref{Q_equal}. So $\{\hat\beta^i\}$ has only a
		single limit point which we denote as $\bar\beta$.

		To prove the second statement, use
		\eqref{bound} to show that
		\begin{align*}
			\frac{\partial}{\partial\beta}
				Q(\beta; \bar\bz,\bar\beta, \bar\btheta, \bar\bgamma)
				\Big|_{\beta=\bar\beta}
			= \frac{\partial}{\partial\beta}
				\call(\bar\bz,\bar\zeta,\beta,\bar\btheta,\bar\bgamma)
				\Big|_{\beta=\bar\beta}.
		\end{align*}
	If $\bar\beta=\varepsilon_\beta$, we have that the derivatives of $Q$ and
	thus of $\call$ are positive. It follows that $\bar\beta$ is a local or
	global minimum. If $\bar\beta\ne\varepsilon_\beta$ it is, by definition, a stationary
	point of $Q$. It is therefore also a stationary point of 
	$\beta\mapsto\call(\bar\bz, \bar\zeta, \beta, \bar\btheta, \bar\bgamma)$. 
\end{IEEEproof}

We can now give a proof of the main convergence result.
\begin{IEEEproof}[Proof of Theorem \ref{thm:convergence}]
	Convergence to a unique limit follows immediately from the assumption and
	Lemmas \ref{lem:z}, \ref{lem:zeta} and \ref{lem:beta}.

	To prove the second statement, we first note that $\call$ is constant with
	respect to those entries of $\btheta$ and $\bgamma$ for which
	$\bar z_k=0$. It then follows from the assumption that
	$\frac{\partial\call}{\partial\theta_k}=0$ and
	$\frac{\partial\call}{\partial\gamma_k}=0$ for all $k=1,\ldots,K_|max|$ at
	the limit point. Similarly from Lemma \ref{lem:zeta} we have
	$\frac{\partial\call}{\partial\zeta}=0$ at the limit point.
	If $\bar\beta\ne\varepsilon_\beta$ we have 
	$\frac{\partial\call}{\partial\beta}=0$ at the limit point and the result
	follows immediately.

	If $\bar\beta=\varepsilon_\beta$ the result can be obtained by introducing
	a Lagrange multiplier such that the limit point satisfies the Karush-Kuhn-Tucker
	conditions.
\end{IEEEproof}

\section{Efficient Evaluation of \texorpdfstring{$\omega_s(i)$, $\omega_t(i)$,
$\omega_v(i)$ and $\omega_x(i)$}{omega s, omega t, omega v and omega x}}
\label{app:superfast}
We derive a low-complexity computation of $\omega_s(i)$
by first inserting \eqref{gohberg-semencul} into \eqref{omega_s} to get
\begin{multline}
	\omega_s(i) = \delta_{N-1}\ii \Bigg( \sum_{q=\max(0,-i)}^{\min(N-1,N-1-i)} \sum_{r=0}^q
		- \rho_{q-r-1} \rho^*_{q+i-r-1} \\
		+ \rho^*_{N-1+r-q} \rho_{N-1+r-q-i}
		\Bigg)
\end{multline}
for $i=-(N-1),\dots,N-1$. Then note that since $\bC$ is Hermitian we have
$\omega_s(-i)=\omega_s^*(i)$. We therefore restrict our attention to
$i\ge0$ in the following. We need the identity
\begin{align}
	\sum_{q=0}^{N-1} \sum_{r=0}^q z_{q,r} = \sum_{q=0}^{N-1} \sum_{k=0}^{N-1-q}
	z_{q+k,k},
	\label{diagsumidentity}
\end{align}
from which we get (recall that $\rho_i=0$ for $i<0$ and $i>N-1$)
{\small
\begin{align*}
	\omega_s(i) = \delta_{N-1}\ii \sum_{q=0}^{N-1-i} (N-i-q)
	(\rho^*_{N-1-q} \rho_{N-1-q-i}
		- \rho_{q-1} \rho^*_{q-1+i}).
\end{align*} }%
Substituting $q=N-1-\bar q-i$ in the first term and $q=\bar q+1$ in the second
term we finally obtain
\begin{align}
	\omega_s(i) = \delta_{N-1}\ii \sum_{\bar q=0}^{N-1} 
	c_s(\bar q, i)
	\rho_{\bar q} \rho^*_{\bar q + i},
\end{align}
where $c_s(\bar q, i) \triangleq (2-N+i+2\bar q)$. The above expression can be
calculated as the sum of two cross-correlations in time $\calo(N\log N)$ by
using FFT techniques.

To evaluate $\omega_t(i)$ \eqref{omega_t} we again insert
\eqref{gohberg-semencul} and get
\begin{multline}
	\omega_t(i) = \delta_{N-1}\ii \Bigg(
	\sum_{q=\max(0,-i)}^{\min(N-1,N-1-i)} 2\pi q
	\sum_{r=0}^q
		- \rho_{q-r-1} \rho^*_{q+i-r-1} \\
		+ \rho^*_{N-1+r-q} \rho_{N-1+r-q-i}
		\Bigg)
\end{multline}
for $i=-(N-1),\dots,N-1$. Be aware that we do not have
$\omega_t(i)=\omega_t^*(-i)$. Applying \eqref{diagsumidentity}, performing the
same substitutions as above and following tedious, but straight-forward,
algebra we finally get
\begin{align}
	\omega_t(i) = \frac{2\pi}{\delta_{N-1}} \sum_{\bar q=0}^{N-1} c_t(\bar q,i)
	\rho_{\bar q} \rho_{\bar q + i}^*
\end{align}
with
\begin{multline}
	c_t(\bar q, i) \triangleq -\bar q(\bar q+i) + i\left( N-\frac{3+i}{2} \right)
	\\
		+ \bar q^2 + (N-1)\left( \bar q - \frac{N-2}{2} \right),
\end{multline}
which again can be evaluated using FFT techniques.

Omitting details, we use a similar approach to find
\begin{align}
	\omega_v(i) &= \frac{4\pi^2}{\delta_{N-1}} \sum_{\bar q=0}^{N-1} c_v(\bar q,i)
		\rho_{\bar q} \rho_{\bar q + i}^* \\
	\omega_x(i) &= \frac{4\pi^2}{\delta_{N-1}} \sum_{\bar q=0}^{N-1} c_x(\bar q,i)
		\rho_{\bar q} \rho_{\bar q + i}^*,
\end{align}
where $\omega_v(-i)= \omega_v^*(i)$. The expression giving $\omega_v(i)$ is
valid for $i\ge 0$, while that giving $\omega_x(i)$ is valid for
$i=-(N-1),\ldots,N-1$. We have also defined
\begin{multline*}
	c_v(\bar q, i)
		= \bar q(\bar q + i)^2
		+ (3\bar q - 2N\bar q - \bar q^2)(\bar q + i) \\
		+ \frac{2}{3}\bar q^3
		+ (N-1)\bar q^2
		+ \left(N^2-3N+\frac{7}{3}\right)\bar q \\
		+ \frac{3}{2}(i-N)^2
		+ \frac{1}{3}(i^3-N^3)
		+ \left(\frac{13}{6} - Mi\right)(i-N) + 1
\end{multline*}%
\vspace{-6mm}
\begin{multline*}
	c_x(\bar q, i)
		= (\bar q^2 + 2\bar q - N\bar q)(\bar q + i) \\
		- \frac{1}{3}\bar q^3
		+ \left(N^2-3N+\frac{7}{3}\right)\bar q
		- \frac{1}{6}i^3 \\
		+ \left(\frac{3N^2-9N+7}{6}\right)i
		- \frac{1}{3}N^3 + \frac{3}{2}N^2 - \frac{13}{6}N + 1.
\end{multline*}

\bibliography{latex/IEEEabrv,latex/refs}

\vfill\break

\begin{IEEEbiography}
[{\includegraphics[width=1in,height=1.25in,clip,keepaspectratio]{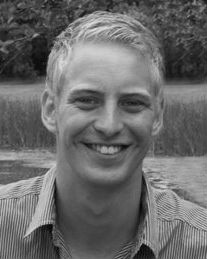}}]
{Thomas Lundgaard Hansen}
received the B.Sc. and M.Sc. (cum laude) in electrical
engineering from Aalborg University, Denmark in 2011 and 2014, respectively.
Since 2014 he has been pursuing the Ph.D. degree in wireless communication at
Aalborg University. During 2013 and 2015 he was a visiting scholar at
the University of California, San Diego, USA. He is the recipient of the best student
paper award (1st place) at the 2014 IEEE Sensor Array and Multichannel Signal Processing
workshop and also received an award from IDA Efondet for his master's thesis.
His research interests include signal processing, machine learning,
optimization and wireless communication.
\end{IEEEbiography}

\begin{IEEEbiography}
[{\includegraphics[width=1in,height=1.25in,clip,keepaspectratio]{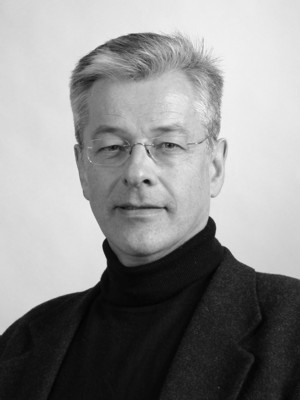}}]
{Bernard Henri Fleury}
(M’97–SM’99) received the Diplomas in Electrical Engineering and in Mathematics in 1978 and 1990 respectively and the Ph.D. Degree in Electrical Engineering in 1990 from the Swiss Federal Institute of Technology Zurich (ETHZ), Switzerland.
Since 1997, he has been with the Department of Electronic Systems, Aalborg University, Denmark, as a Professor of Communication Theory. From 2000 till 2014 he was Head of Section, first of the Digital Signal Processing Section and later of the Navigation and Communications Section. From 2006 to 2009, he was partly affiliated as a Key Researcher with the Telecommunications Research Center Vienna (ftw.), Austria.
Prof. Fleury’s research interests cover numerous aspects within communication theory, signal processing, and machine learning, mainly for wireless communication systems and networks.

\end{IEEEbiography}

\begin{IEEEbiography}
[{\includegraphics[width=1in,height=1.25in,clip,keepaspectratio]{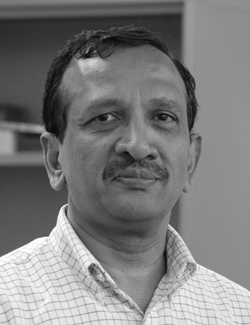}}]
{Bhaskar D. Rao}
(S’80–M’83–SM’91–F’00)
is currently a Distinguished Professor in the Electrical and Computer Engineering department and the holder of the Ericsson endowed chair in Wireless Access Networks at the University of California, San Diego. Prof. Rao was elected fellow of IEEE in 2000 and is the recipient of the 2016 IEEE Signal Processing Society Technical Achievement Award. Prof. Rao’s interests are in the areas of digital signal processing, estimation theory, and optimization theory, with applications to digital communications, speech signal processing, and biomedical signal processing.
\end{IEEEbiography}

\vfill

\end{document}